\documentclass[journal]{IEEEtran}
\IEEEoverridecommandlockouts  

\usepackage{generic}
\usepackage{cite}
\usepackage{amsmath}
\usepackage{amssymb}
\usepackage{algorithmic}
\usepackage{graphicx}
\usepackage{textcomp}
\usepackage{dsfont}
\usepackage{breqn} 
\usepackage{algorithm}
\usepackage{algorithmic}
\usepackage{nicefrac}       
\usepackage{microtype}      
\usepackage{verbatim}
\usepackage{flushend}
\allowdisplaybreaks
\usepackage{xcolor}
\usepackage{soul}

\usepackage{microtype}      
\usepackage{verbatim}
\usepackage{flushend}
\allowdisplaybreaks
\usepackage{soul}

\newcommand{\conv}{\text{conv}}

\newtheorem{theorem}{Theorem}[section]

\newtheorem{lemma}[theorem]{Lemma}
\newtheorem{definition}{Definition}[section]

\newtheorem{thm}{Theorem}

\newtheorem{remark}[thm]{Remark}
\newtheorem{assumption}{Assumption}[section]
\newtheorem{proof}{Proof}[section]
\usepackage{algorithm}
\usepackage{algorithmic}

\usepackage{verbatim}
\def\BibTex{{\rm B\kern-.05em{\sc i\kern-.025em b}\kern-.08em
    T\kern-.1667em\lower.7ex\hbox{E}\kern-.125emx}}
\begin{document}
\title{Safe Learning MPC with Limited \\Model Knowledge and Data} 


\author{Aaron Kandel and Scott J. Moura
\thanks{Submitted for review on June 10th, 2021. This work was supported by a National Science Foundation Graduate Research Fellowship.
}
\thanks{Aaron Kandel is affiliated with the Department of Mechanical Engineering at the University of California, Berkeley, Berkeley, CA 94709 USA (e-mail: aaronkandel@berkeley.edu).}
\thanks{Scott Moura is affiliated with 
the Department of Civil and Environmental Engineering at the University of California, Berkeley, Berkeley, CA 94709 USA (e-mail:  smoura@berkeley.edu).}}

\maketitle

\begin{abstract} 

This paper presents an end-to-end framework for safe learning-based control (LbC) using nonlinear stochastic MPC and distributionally robust optimization (DRO). This work is motivated by several open challenges in LbC literature. In particular, many control-theoretic LbC methods require subject matter expertise in order to translate their own safety guarantees, often manifested as preexisting data of safe trajectories or structural model knowledge.  In this paper, we focus on LbC where the controller is applied directly to a system of which it has no or extremely limited direct experience, towards safety during \textit{tabula-rasa} or ``\textit{blank slate''} model-based learning and control as a challenging case for validation. This explores the boundary of the status-quo in control theory relating to requirements for subject matter expertise.  We show under basic and limited assumptions on the underlying problem, we can translate probabilistic guarantees on feasibility to nonlinear systems using results in stochastic MPC and DRO literature whose relevance we formally extend in a mathematical analysis. We also present a coupled and intuitive formulation for persistence of excitation (PoE), and illustrate the connection between PoE and applicability of the proposed method.  Our case studies of vehicle obstacle avoidance and safe extreme fast charging of lithium-ion batteries reveal powerful empirical results supporting the underlying DRO theory. Our method is widely applicable within the LbC domain to, for example, airborne wind energy systems, vehicle obstacle avoidance, and energy storage systems management. It is also applicable to quantifying uncertainty beyond the LbC case. 

\end{abstract}

\begin{IEEEkeywords}
learning, adaptive control, data-driven control, robust optimization, model-predictive control, energy systems, lithium-ion battery, vehicle autonomy
\end{IEEEkeywords}

\section{Introduction}
\label{sec:introduction}
\IEEEPARstart{T}{his} paper presents a novel application of Wasserstein ambiguity sets to robustify model-based reinforcement learning (MBRL) and learning-based control (LbC) in safety-critical applications. Here, we define safety as the ability of the control policy to satisfy constraints. Translating safety to online reinforcement learning (RL) algorithms is a notoriously difficult open challenge in relevant literature. This paper is motivated by unsolved shortcomings of many existing means to address this challenge, particularly a strong and often optimistic dependence on subject matter expertise.  Two overarching examples include (i) assumed knowledge of underlying dynamics, and (ii) preexisting data of safe trajectories.

The LbC problem space borrows many concepts from historical research on stochastic optimal control, a field which dates back decades to the original linear-quadratic Gaussian problem \cite{Karl00}.  The key underlying concept relates to uncertainty, and how we can accommodate limited or imperfect knowledge of the underlying dynamics. The rise in popularity of MPC has created a new application for these robust and stochastic control principles.  For instance, foundational work by Kothare et al. addresses uncertainty in MPC optimization with linear matrix inequalities by allowing the state transition matrices to vary in time within a convex polytope \cite{Kothare00}. 

Within the past few years, stochastic optimal control has become connected to ongoing research in the burgeoning field of LbC.  Here, researchers seek guarantees on safety and performance when learning-based controlling a dynamical system simultaneously. For a review of current state of the art methods in learning-based control which utilize MPC, we direct the reader to a thorough review by Hewing et al. \cite{Hewing00}.  This type of problem presents a nuanced and complex challenge for a host of reasons.  Safety and feasibility pose significant barriers for proper implementation of such algorithms.  Moreover, balancing the exploration-exploitation tradeoff inherent to simultaneous control and model identification has presented researchers with a host of unique problems which form a primary focus of research in active learning.  Work by Dean et al., for instance, explores safety and persistence of excitation for a learned constrained linear-quadratic regulator \cite{Dean00}. 

MPC is a highly popular use case for learning-based control problems, and provides an intuitive bridge between longstanding adaptive control theory and new developments and explorations.  For instance, recent work has investigated recursive feasibility for adaptive MPC controllers based on recursive least-squares \cite{bujarbaruah2018adaptive} and set-membership parameter identification \cite{Tanaskovic00}, although similar papers frequently possess limitations including a dependence on linear dynamical models.   Rosolia and Borrelli derive recursive feasibility and performance guarantees for a learned episodic MPC controller \cite{Rosolia00}.  Koller et al. also address the safety of a learned MPC controller when imperfect model knowledge and safe control exists \cite{Koller00}.

We note that Control Lyapunov function and control barrier function \cite{Cheng2019EndtoEndSR,Fan2020BayesianLA,Choi00} based approaches have further strengthened the connection between classical adaptive control and more modern approaches akin to popular model-based reinforcement learning (RL) problems.  Recent work by Westenbrouk et al. has even explored coupling such nonlinear control methods with a policy optimization scheme \cite{Westenbroek00}.

In the space of RL, safe LbC has become a burgeoning area of study. For broad discussion and categorization of classical methods, Garcia et al. provide a comprehensive review \cite{Garcia00}. More recently, some control-theoretic principles have migrated towards the space of safe RL. For example, Chow and Nachum leverage Lyapunov stability principles to obtain improved empirical results \cite{NEURIPS2018_4fe51490}. Other methods focus on safety as a challenge relevant to transfer learning, where safe behavior can be extrapolated and expanded from simpler tasks \cite{safcrit00}. Methods in the space of RL provide idealistic safety guarantees that translate into improved empirical safety properties. However, any guarantees (probabilistic or robust) or safety certificates in this space are elusive and remain an open challenge.


Guarantees in RL literature are difficult to obtain since that literature eschews subject matter expertise (SME), or direct intuition into a specific application.  Some RL research obtains guarantees by leveraging strong SME in the form of known safe backup controllers \cite{preexsafe0,preexsafe00}. Generally, when RL neglects considerations to SME it becomes applicable to a much wider body of relevant decision and control problems \cite{dvp00} that lack permeability to our intuition and expertise. Conversely, controls literature is ubiquitous in revealing how such expertise can be leveraged to yield strong and \textit{specific} performance and safety even in adaptive and learning contexts. As previously discussed, SME in controls LbC methods often takes the form of model knowledge \cite{bujarbaruah2018adaptive,Tanaskovic00, Cheng2019EndtoEndSR,Fan2020BayesianLA,Choi00} and preexisting data of safe trajectories \cite{Rosolia00, deepc00}.  

The problem with these SME assumptions is that they can very easily become optimistic.  Given the overarching assumption of preexisting data of safe trajectories, we have to ask ``How trustworthy is our data?'' This should always be called into question, especially when safety is of the utmost importance. Many LbC methods do consider noise-corrupted data \cite{deepc00}, but what if deeper, malicious pathology infiltrates the data generation process? The process generating the data could be flawed in many ways, the relevance of each to existing methods varies but is persistent.  An example could be sampling data locally where relevant dynamics can be effectively linearized, when the system experiences highly nonlinear behavior outside of that region. Without exploiting and trusting our SME, we cannot guarantee things like this will not happen especially in safety-critical settings. By applying a resultant controller to the underlying system, it can encounter out-of-distribution (OOD) experience and adversarial attacks that a majority of existing LbC methods simply cannot accommodate. Those few LbC algorithms that do make consideration to OOD experience do so using hyperparameters that are not trivial to select and validate \cite{deepc00}, and often assume structure of the underlying dynamics \cite{MPC_new00}. These same fundamental quandaries also apply when assuming model knowledge.  

In this paper, we address these key open questions about SME in control theoretic LbC. Critically, we ask ``\textit{What is the least amount of SME we may need to obtain safe control results?}'' Such questions remain relatively unexplored in controls literature, despite their relevance. Our methods for addressing these questions are actually quite simple, and rely on combination of concepts in stochastic MPC and distributionally robust optimization. We make this technical augmentation along with several basic assumptions about the problem formulation that allow us to translate probabilistic safety guarantees in the absence of conventionally strong dependence on SME.

\subsection{Background on DRO and LbC}

This paper primarily leverages concepts from distributionally robust optimization (DRO) to obtain safety certificates. 
In recent practice, DRO has been gaining traction as a set of methods that provide significant value to the study and solution of the LbC problem. DRO is a field of inquiry which seeks to guarantee robust solutions to optimization programs when the distributions of relevant random variables are estimated via sampling.  This uncertainty can involve the objective or the constraints of the optimization program. Uncertainty in both cases can pose significant challenges if unaccounted for, leading to suboptimal and potentially unsafe performance \cite{Nilim00}.  Given that past work in the LbC space frequently considers chance constraints \cite{bujarbaruah2018adaptive, Khojasteh2020ProbabilisticSC, deepc00}, incorporating a true DRO approach possesses the potential to improve our capabilities of guaranteeing safety during learning.  These methods have been recently explored to address challenges of safety and performance imposed by uncertainty. For instance, Van Parys et al. address distributional uncertainty of a random exogenous disturbance process with a moment-based framework \cite{VanParys00}.  Paulson et al. also apply polynomial chaos expansions to characterize distributional parametric uncertainty in a nonlinear model-predictive control application \cite{Paulson00}.   

Within the toolbox provided by DRO, Wasserstein ambiguity sets are a foremost asset.  The Wasserstein metric (or ``earth mover's distance'') is a symmetric distance measure in the space of probability distributions.  Wasserstein ambiguity sets account for distributional uncertainty in a random variable, frequently one approximated in a data-driven application.  They accomplish this feat with out-of-sample performance guarantees by repLbCing the data-driven distribution of the random variable with the worst-case realization within a Wasserstein ball centered about the empirical distribution \cite{Esfahani00, Gao00}.  Expressions exist which map the quality of the empirical distribution with Wasserstein ball radii such that desired robustness characteristics are achieved without significant sacrifices to the performance of the solution \cite{Zhao00}.  Within the control context, however, the Wasserstein distance metric has only recently began emerging as a valuable and widespread tool.  Work by Yang et al. explores the application of Wasserstein ambiguity sets for distributionally robust control subject to disturbance processes \cite{Yang00}.  Similar methods have made their way to research on model-based and model-free reinforcement learning as well \cite{kandel2021distributionally,Kandel01, MPC_new00}. DRO has also been applied to Markov decision processes (MDPs) in a general sense, with good results \cite{NIPS2019_8942, asadi2018lipschitz,Akbar00,Yang03}.  Scalability is still an open challenge in that space.  Overall, while Wasserstein ambiguity sets are seeing increased application in controls research, many of their true capabilities have yet to be fully exploited.

\subsection{Statement of Contributions}

This paper seeks to address key shortcomings in these areas of literature.  Among those previously discussed, foremost is the lack of general methods that possess robustness when conducting \textit{tabula-rasa} learning-based control, or those requiring significant assumptions on availability of prior data of safe control trajectories. 

We present a novel and simple model-based LbC scheme based on MPC which provides strong probabilistic out-of-sample guarantees on safety. We validate our method using experiments that emulate \textit{tabula-rasa} as closely as possible given our assumptions, but our algorithm is widely applicable to adaptive control scenarios especially when underlying dynamics may be poorly structured or difficult to characterize. By developing Wasserstein ambiguity sets relating to empirical distributions of modeling error, we can conduct MPC with an imperfect learned snapshot model while maintaining confidence on our ability to satisfy nominal constraints.  The Wasserstein ambiguity sets allow us to optimize with respect to constraint boundaries that are shifted into the safe region.  As our empirical distributions improve with more online data generation, the offset variables tighten towards the nominal boundary in a provably safe way.  Critically, in this paper, we present this LbC scheme along with (1) an explicit and fundamental persistence of excitation (PoE) scheme, and (2) highly limited SME assumptions. While many LbC methods are amenable to PoE schemes \cite{Dean00}, the question of PoE is in some cases neglected despite its relevance. We actually show our explicit PoE scheme is fundamental to illustrating the applicability of our method. Our contributions combine to allow us to translate safety guarantees with no strong model knowledge or prior data of existing safe trajectories. 

Our approach yields probabilistic safety guarantees. The overarching objective of this paper is not to present the most high-performing LbC architecture, but rather to explore what kind of performance we can obtain when limiting our SME assumptions moreso than existing work in controls literature.  Many control-theoretic methods provide stronger robust (i.e. safety w.p. 1) guarantees under much more restrictive assumptions. In our case, we label our method as ``trustworthy'' insofar as it relies on highly limited SME. Given the elusiveness of safety guarantees in RL literature, a probabilistic result within our context is powerful.



We validate our approach by learning to safely fast charge a lithium-ion battery using a nonlinear equivalent circuit model.  Battery fast charging presents a strong challenge for learning-based control methods, given that the optimal policy is a boundary solution which rides constraints until the terminal conditions are met.  We also conduct a case study on safe autonomous driving using a nonlinear bicycle model of vehicle dynamics.  We demonstrate that our algorithm provides a provably safe method for the vehicle to avoid obstacles while learning its dynamics from scratch. 

We provide an open-source GitHub repository \cite{KandelGitHub} for our case studies.

\section{Distributionally Robust Optimization}
The core of our proposed algorithmic architecture relies heavily on distributionally robust optimization (DRO) techniques.  In the following section, we outline fundamental ideas which establish the foundation of our algorithm.

\subsection{Chance Constrained Programming}
A chance constraint is a constraint within an optimization program which is only satisfied with some probability.  This is typically a necessary concession when the constraint is affected by a random variable $\bf{R}$:
\begin{equation}\label{eqn:cc1}
    {\mathbb{P}} \big{[}h(x_k, u_k, \textbf{R}) \leq 0\big{]} \geq 1 - \eta 
\end{equation}
Here, the constraint function $h(x_k,u_k,\textbf{R})$ outputs an $m$-dimensional vector.
In this case, the distribution $\mathbb{P}$ relates to random variable $\bf{R}$ with support $\xi$. Here, $0 \leq \eta < 1$ is the specified risk metric or our allowed probability to violate the constraint. If $\eta=0$, we say we have a robust optimization program which must not yield \textit{any} probability of constraint violation.  In practice, especially when approximating $\mathbb{P}$ from sampling, we admit some small probability of constraint violation leading to a value of $\eta>0$.  This is frequently necessary because it allows our probabilistically robust solution to balance conservatism with performance.  

Upon utilizing an empirical approximation of $\mathbb{P}$ derived from sampling (usually denoted $\hat{\mathbb{P}}$), we admit some distributional uncertainty which can arise from only having access to a finite group of samples.  The law of large numbers states that for any number of samples $\ell \rightarrow \infty$, $\hat{\mathbb{P}}\rightarrow \mathbb{P}^*$. The discrepancy from this limited sampling creates distributional uncertainty, which can affect the quality of the solution if our approximation $\hat{\mathbb{P}}$ is inaccurate \cite{Nilim00}.  Throughout the remainder of this section, we discuss the application of distributionally robust optimization techniques to address this distributional uncertainty.

\subsection{Wasserstein Ambiguity Sets}
The Wasserstein metric is defined as follows:
\begin{definition}
Given two marginal probability distributions $\mathbb{P}_1$ and $\mathbb{P}_2$ lying within the set of feasible probability distributions $\mathcal{P}(\xi)$, the Wasserstein distance between them is defined by
\begin{equation}
    \mathcal{W}(\mathbb{P}_1, \mathbb{P}_2) = \underset{\Pi}{\text{inf}} \bigg{\{} \int_{\xi^2} ||\textbf{R}_1 - \textbf{R}_2 ||_a \Pi (d\textbf{R}_1, d\textbf{R}_2) \bigg{\}}
\end{equation}
where $\Pi$ is a joint distribution of the random variables $\bf{R}_1$ and $\bf{R}_2$, and $a$ denotes any norm in $\mathbb{R}^n$. 
\end{definition}

The Wasserstein metric is colloquially referred to as the ``earth-movers distance.'' This name is rooted in the interpretation of the Wasserstein metric as the minimum cost of redistributing mass from one distribution to another via non-uniform perturbation \cite{Yang00}. To show why the Wasserstein distance is a valuable tool we can leverage to robustify a data-driven optimization program, we first reference the chance constraint equation (\ref{eqn:cc1}), which depends on an empirical distribution $\hat{\mathbb{P}}$. Rather than solving the optimization program with respect to an imperfect snapshot of $\mathbb{P}^*$ defined by $\hat{\mathbb{P}}$, we can optimize over any probability distribution within some ambiguity set centered around our estimate $\hat{\mathbb{P}}$. The Wasserstein distance provides a formal method to define such an ambiguity set.  Namely, we can optimize against the worst-case realization of $\textbf{R}$ sourced from a set of probability distributions within specified Wasserstein radius of our empirical estimate. We define ``worst-case'' as the realization which yields the lowest probability of satisfying the chance constraint. This formulation can be described mathematically with the following relation:
\begin{equation}\label{eqn:wass3}
    \underset{\mathbb{P} \in \mathbb{B}_\epsilon}{\text{inf}} \;  \mathbb{P} \big{[} h(x_k, u_k, \textbf{R}) \leq 0 \big{]} \geq 1 - \eta 
\end{equation}
where
\begin{equation}\label{eqn:wass1}
    \mathbb{B}_\epsilon := \big{\{} \mathbb{P} \in \mathcal{P}(\xi) \; | \; \mathcal{W}(\mathbb{P}, \hat{\mathbb{P}}) \leq \epsilon \big{\}}
\end{equation}
is the ambiguity set defined for a Wasserstein ball radius $\epsilon$. Of note is the fact that (\ref{eqn:wass3}) guarantees probabilistic feasibility for any probability distribution within the ambiguity set when reformulated correctly.  No assumptions must be leveled on the true distribution $\mathbb{P}^*$ for these guarantees to translate under a proper reformulation.  

Reformulation is necessary because the exact constraint shown in (\ref{eqn:wass3}) poses an infinite dimensional nonconvex problem.   Ongoing research has pursued tractable reformulations of this constraint which facilitate its real-time solution.

This paper adopts a reformulation of (\ref{eqn:wass3}) detailed in \cite{Duan00}. This reformulation accommodates vector constraint functions and requires that the function $g(x_{k}, u_{k}, \bf{R})$ is linear in $\textbf{R}$, and entails a scalar convex optimization program to derive. Our algorithm is designed to exploit the linear dependence on R such that this assumption has no affect on the applicability of our approach. Importantly, the result is a conservative \textit{convexity-preserving} approximation of (\ref{eqn:wass3}).  For an $m$-dimensional constraint function, the exact form of the ambiguity set is $\mathcal{V} = \conv(\{r^{(1)}, ..., r^{(2^m)}\})$, where the vector $r$ is sourced from the optimization component of the overall procedure.  The set of constraints we find to replace the infinite dimensional DRO chance constraint are:
\begin{align}
    &h(x_{k},u_{k}) + r^{(j)} \leq 0,  &\forall \ j=1,...,2^m \label{eqn:ineq-reform}
\end{align}
For complete and elegant discussion of this reformulation, we highly recommend the reader reference work in \cite{Duan00}, specifically pages 5-7 of their paper. This reformulation requires some additional information, including a tractable representation of an appropriate Wasserstein ball radius.



Finally, several expressions exist for the Wasserstein ball radius $\epsilon$ which are probabilistically guaranteed to contain the true distribution with allowed probability $\beta$. We adopt the following formulation of $\epsilon$ from \cite{Zhao00} \begin{equation}\label{eqn:wass2}
    \epsilon(\ell) = C \sqrt{\frac{2}{\ell} \log \bigg{(} \frac{1}{1-\beta} \bigg{)} }
\end{equation}
where $\ell$ is the number of data points, $\beta$ is the probability the Wasserstein ball contains the true distribution, and $C$ relates to the diameter of the support of the distribution and is obtained by solving the following scalar optimization program:
\begin{equation}\label{eqn:wass4}
    C \approx 2 \: \underset{\alpha > 0} {\text{inf}} \left\{ \frac{1}{2\alpha}\left( 1 + \ln \left(\frac{1}{\ell}\sum_{k=1}^N e^{\alpha ||\textbf{R}^{_k}-\hat{\mu}||_1^2}\right)\right)\right\}^{\frac{1}{2}}
\end{equation}
where the right side bounds the value of $C$, and $\textbf{R}^{_k}$ is a sample of the random variable which comprises our empirical distribution, and $\bar{\mu}$ is the sample mean of the distribution.

\section{Equivalent Chance-Constraint Reformulation}  

This paper builds upon the equivalent reformulation of (\ref{eqn:wass3}) from \cite{Duan00}. This reformulation leverages findings from recent work by \cite{Esfahani00}. The statement of the specific reformulation in \cite{Duan00} indicates a requirement that the constraint function $g(x, \bf{R})$ is linear in $x$ and $R$, respectively. 

Notably, we identify a simple extension of the reformulation in \cite{Duan00} that allows its application to our nonlinear MPC formulation via relaxing requirement the constraint function be linear in the decision variable $x$.

\subsection{Restatement of the Reformulation from \cite{Duan00}}

The reformulation from \cite{Duan00} is stated to require the constraint function $g(x, \bf{R})$ to be linear in $x$ and $\textbf{R}$, respectively. In the next subsection, we extend the reformulation to include some broader cases of constraint functions:
\begin{equation}
    g(x, \textbf{R}) = g_x(x) + g_R(\textbf{R}).
\end{equation}
where the functions $g_x$ and $g_R$ can be nonlinear in their respective arguments.
In this subsection, we restate the work from \cite{Duan00} as a reference for our extension included in subsection III.b.

Data samples $\{R^{(1)}, R^{(2)}, ..., R^{(\ell)} \}$ corresponding to random variable $\bf{R} \in \mathbb{R}^m$ are drawn from the true distribution $\mathbb{P}^*$. These finite samples comprise our empirical distribution $\hat{\mathbb{P}}$. The finite-ness of our empirical distribution indicates it will not perfectly match the behavior of the true distribution $\mathbb{P}^*$. This is especially true in cases with limited samles, which are relevant to the challenging case studies this paper explores.  

Normalizing the data lends simplicity to the derivation: 
\begin{equation}
    \vartheta^{(i)} = \Sigma^{-\frac{1}{2}}({R}^{(i)}-\mu)
\end{equation}
where $\Sigma$ is the sample variance of the data and $\mu$ is the sample mean. This standardization transforms the data samples such that its new mean is $0$, and its new variance is $I_{m\times m}$. The support of this normalized distribution is
\begin{equation}
    \Theta = \{\vartheta \in \mathbb{R}^m \ | \ -\sigma_{\max} \textbf{1}_m \leq \vartheta \leq \sigma_{\max} \textbf{1}_m\}
\end{equation}
since we have centered the normalized variable $\vartheta$. Note that $\bf{1}_m$ is a column vector of ones.  Let $\mathbb{Q}^*$ and $\hat{\mathbb{Q}}$ represent the true and empirical distributions of the normalized data $\vartheta$. We construct the ambiguity set ${\hat{\mathcal{Q}}}$ using the ``Wasserstein ball'' given by (\ref{eqn:wass1}), allowing us to transform the distributionally robust chance constraint (DRCC) in \eqref{eqn:wass3} to  
\begin{equation}
    \underset{\mathbb{Q} \in \hat{\mathcal{Q}}}{\text{sup}} \mathbb{Q}[ \vartheta \notin \mathcal{V}] \leq \eta
\end{equation}
which says the worst case probability that normalized random variable $\vartheta$ is outside set $\mathcal{V}$ is less than $\eta$, where the supremum is taken over all distributions $\mathcal{Q}$ in ambiguity set $\hat{\mathcal{{Q}}}$.
We wish to obtain the least conservative (i.e. tightest) set $\mathcal{V} \subseteq \mathbb{R}^m$ in order to define the desired Wasserstein uncertainty set $\mathcal{A} = \left\{ a \in \mathbb{R}^m \ | \ a = \Sigma^{\frac{1}{2}} v + \mu, \ v \in \mathcal{V} \right\}$ such that
\begin{equation}
    g(x_k, u_k, \bf{R}) \leq 0, \; \forall \; \bf{R} \in \mathcal{A}
\end{equation}
We restrict the overall shape of the set $\mathcal{V}$ to be a hypercube, which enables computational tractability:
\begin{equation}
    \mathcal{V}(\sigma) = \{ \vartheta \in \mathbb{R}^m | -\sigma \boldmath{1}_m < \vartheta < \sigma \boldmath{1}_m \}.
\end{equation}
Now, to compute this ambiguity set without introducing unnecessary conservatism, we need to find the minimum value of the hypercube side length $\sigma \in \mathbb{R}$.  The following optimization program details this problem:
\newpage
\begin{align}
\underset{0\leq\sigma\leq \hat{\sigma}_{max}} {\text{min}}  &  \sigma \\ %
\text{subject to:} \quad &
\underset{\mathbb{Q} \in \hat{\mathcal{Q}}} {\text{sup}} \: \mathbb{Q}[\tilde{\vartheta} \notin \mathcal{V}(\sigma)]\leq \eta
\end{align}
Here, we select $\hat{\sigma}_{max}$ using \textit{a priori} information about the specific problem context.

The derivation in \cite{Duan00} provides a worst-case probability formulation, summarized by the following Lemma:
\begin{lemma}[Lemma 2 of \cite{Duan00}]
\begin{equation}
\begin{aligned} 
\underset{\mathbb{Q} \in \hat{\mathcal{Q}}} {\text{sup}} \mathbb{Q}[\tilde{\vartheta} \notin \mathcal{V}(\sigma)]=\\
\underset{\lambda \geq 0} {\text{inf}} \bigg{\{} \lambda \epsilon(\ell) + \frac{1}{\ell} \sum_{j=1}^\ell \left(1-\lambda \left(\sigma- ||\vartheta^{(j)}||_\infty \right)^+\right)^+\bigg{\}} \label{eqn:wass4}
\end{aligned}
\end{equation}
where $(x)^+=\max(x,0)$.
\end{lemma}

We defer to \cite{Duan00} for the proof of this finding. Their result entails that (\ref{eqn:wass4}) can be reformulated as
\begin{equation}
     \underset{0 \leq \lambda,0\leq\sigma\leq \hat{\sigma}_{max}}  {\text{min}} \sigma \qquad \text{subject to:} \quad h(\sigma, \lambda) \leq \eta \label{sigeqn} \leq\sigma_{max}
\end{equation}
where 
\begin{equation}
    h(\sigma, \lambda) = \lambda \epsilon(\ell) + \frac{1}{\ell} \sum_{j=1}^\ell \left( 1-\lambda(\sigma- ||\vartheta^{(j)}||_\infty)^+\right)^+
\end{equation}
The result of this optimization program is the value of $\sigma$, which is used to reformulate the chance constraints via convex approximation.  For a convex approximation of the constraint function in (\ref{eqn:wass3}), the hypercube $\mathcal{V}(\sigma)$ becomes the convex hull of its vertices.  If for example $m=1$ (i.e. the random variable is 1-dimensional), then $\mathcal{V}(\sigma)=(-\sigma, \sigma)$ -- an open interval. The offset $r^{(j)}$ is calculated from:
\begin{align}
    r^{(1)}&=\Sigma^{\frac{1}{2}}\textbf{1}_m\sigma+\mu\\
    r^{(2)}&=\Sigma^{\frac{1}{2}}\textbf{1}_m(-\sigma)+\mu
\end{align}
In the two dimensional case, this yields the ambiguity set $\mathcal{A} = \text{conv}(\{\pm \sigma, \pm \sigma\})$ where $\text{conv}(\{\cdots\})$ represents the convex hull of points $\{\cdots\}$.  For an $m$-dimensional constraint function, the exact form of the reformulated ambiguity set is $\mathcal{V} = \text{conv}(\{r^{(1)}, ..., r^{(2^m)}\})$. In each case, the ambiguity set is a hypercube, and the change of signs is the method by which we enumerate across that hypercube's vertices. The set of constraints are:
\begin{align}
    &g(x) + r^{(j)} \leq 0,  &\forall \ j=1,...,2^m
\end{align}
Algorithm 1 details the method used to compute the offset $\sigma$. 
\begin{algorithm}[tb]
   \caption{Computation of $\sigma$}
   \label{alg:example}
\begin{algorithmic}
\item Initialize $\underline{\sigma}=0, \bar{\sigma}=\sigma_{max}$
\WHILE{$\bar{\sigma}-\underline{\sigma}>\text{tolerance}$}
   \item $\sigma = \frac{\bar{\sigma}+\underline{\sigma}}{2}$
   \item $[\lambda,h^*(\sigma,\lambda)]$ = minimize($\sigma$, $\lambda_{lb}$, $\lambda_{ub}$, $\epsilon$, $\theta$)
   \IF{$h^*(\sigma,\lambda)>\eta$}
   \item $\underline{\sigma}=\sigma$
   \ELSE
   \item $\bar{\sigma}=\sigma$
   \ENDIF
   \ENDWHILE
   \item $\sigma=\bar{\sigma}$
\end{algorithmic}
\end{algorithm}

\subsection{Extending the Reformulation}
Duan et al. utilize the findings of \cite{Esfahani00} in presenting their convex reformulation. Critically, we identify that the fundamental theory presented by \cite{Esfahani00} allows applying the identical reformulation to cases where the constraint function takes the form
\begin{equation}
    g(x, \textbf{R}) = g_x(x) + g_R(\textbf{R}).
\end{equation}
wherein $g_x$ and $g_R$ may be nonlinear functions.
Critically, there must not be any interdependence between $x$ and $\textbf{R}$.

This paper presents a modified lemma for the applicability of the previously stated reformulation first presented by \cite{Duan00}. 
\begin{lemma}
    If the function $g$ satisfies 
    \begin{equation}
        g(x, \textbf{R}) = g_x(x) + g_R(\textbf{R}). \label{newg}
    \end{equation}
    then constraints of the following form: 
    \begin{equation}
        \underset{\mathbb{P} \in \mathbb{B}_\epsilon}{\text{inf}} \;  \mathbb{P} \big{[} g(x, \textbf{R}) \leq 0 \big{]} \geq 1 - \eta \label{DROCC}
    \end{equation}
    can be reformulated into the convex approximation
    \begin{align}
        &g_x(x) + r^{(j)} \leq 0,  &\forall \ j=1,...,2^m
    \end{align} 
    using the relations in (\ref{eqn:wass4}-\ref{sigeqn}),
    where $r=\Sigma^{\frac{1}{2}}\textbf{1}_m\sigma+\mu$. 

\end{lemma}
\begin{proof}
We start by defining auxiliary variables in the constraint function. Consider that, without loss of generality, nonlinear functions of $\textbf{R}$ can themselves be considered the random variable in question:
    \begin{equation}
        \tilde{\textbf{R}} = g_R(\textbf{R})
    \end{equation}
    where $\tilde{\textbf{R}}$ is the new model of the stochasticity. This gives
    \begin{equation}
        g(x,\textbf{R}) = g_x(x) + \tilde{\textbf{R}}
    \end{equation}
Now, we create a dummy auxiliary decision variable $\tilde{x}$ in the same manner: 
\begin{equation}
    \tilde{g}(\tilde{x}, \tilde{\textbf{R}}) = \tilde{x} + \tilde{\textbf{R}}
\end{equation}
forming a function $\tilde{g}$ which is trivially linear in $\tilde{x}$ and $\tilde{\textbf{R}}$, where
\begin{align}
    \tilde{x} &= g_x(x). \label{eqcon} 
\end{align}
This equality constraint (\ref{eqcon}) now shows up in the overall optimization program. However, the DRCC reformulation only poses conditions on the constraint function in question (namely $\tilde{g}(\tilde{x}, \tilde{\textbf{R}})$).
We have transformed the distributionally robust chance constraint into
\begin{equation}
     \underset{\mathbb{P} \in \mathbb{B}_\epsilon} {\text{inf}} \mathbb{P} \left[ \tilde{g}(\tilde{x}, \tilde{\textbf{R}}) \leq 0 \right] \geq 1 - \eta
\end{equation}
which is now linear in $\tilde{x}$ and $\tilde{\textbf{R}}$.
Following procedure from \cite{Esfahani00}, we suppress dependence on $x$ (or $\tilde{x}$) for simplicity, leading to $\ell(\tilde{\textbf{R}})=\tilde{g}(\tilde{x}, \tilde{\textbf{R}})$ \cite{Esfahani00, Duan00}:
\begin{equation}
     \underset{\mathbb{P} \in \mathbb{B}_\epsilon} {\text{inf}} \mathbb{P} \left[ \ell(\tilde{\textbf{R}}) \leq 0 \right] \geq 1 - \eta. \label{drccsup}
\end{equation}
The remainder of the proof is identical to the Appendix in \cite{Duan00}, leading to the convex approximation:
\begin{align}
        &g_x(x) + r^{(j)} \leq 0,  &\forall \ j=1,...,2^m
    \end{align}

Beyond exploiting the linear presence of $\tilde{x}$ in the constraint function, suppressing dependence on decision variables is possible and helpful for the following reasons. The overall process of solving an optimization program with a DRCC is characterized by a two stage stochastic optimization problem. Here, (\ref{drccsup}) is the first stage problem that we solve using the equivalent reformulation. Esfahani and Kuhn show in Section 5.3 of their paper that, without loss of generality, the solution in the second stage (i.e. the overall optimization program) is unaffected by suppressing dependence of $\ell$ on decision variables in the first stage.  Additionally, the decision-independent loss function $\ell(\tilde{\textbf{R}})$ can trivially be expressed as a pointwise maximum of elementary measurable functions, as required by Section 4 of \cite{Esfahani00}. 

This means that, in practice, the dummy decision variable $\tilde{x}$ will not come into play during any stage of solution. After solving the first stage problem, we can reverse the substitution in the remaining optimization to avoid an equality constraint with poor computational tractability. 
\end{proof}

We have shown a simple extension of the DRO reformulation from \cite{Duan00} that allows us to apply the method to nonlinear optimization programs. In the next section of this paper, we describe our nonlinear MPC formulation and the context within which the guarantee from the DRCC is translated to LbC. 


\section{Distributionally Robust Model-Based learning-based control}
\begin{figure*}[ht!]\label{fig:res1}
      \centering  
      \includegraphics[trim = 0mm 0mm 0mm 0mm, clip, width=\textwidth]{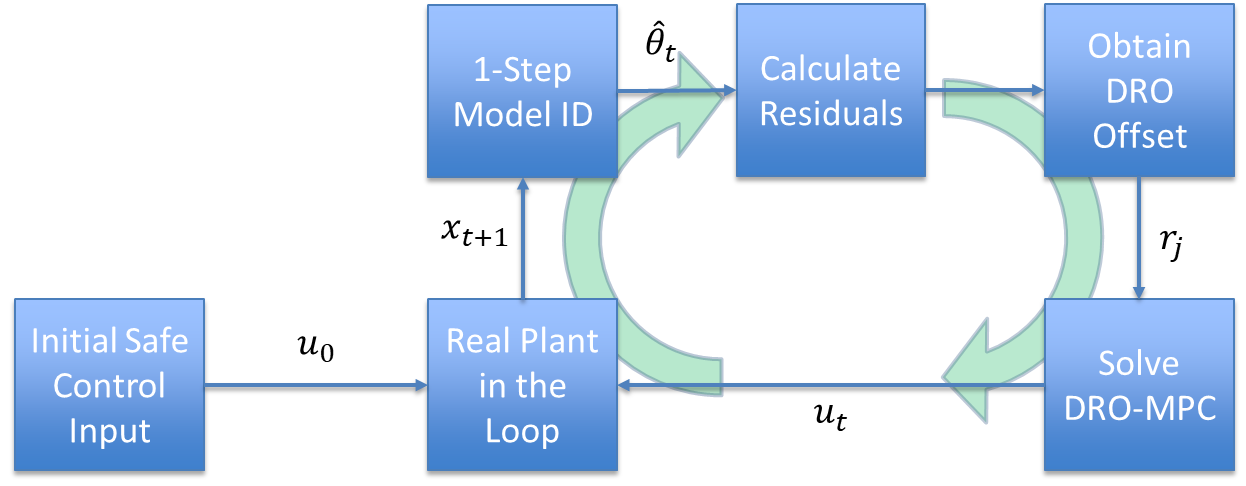}
      \caption{Diagram of safe Wasserstein-constrained MPC. In the most restrictive case, after initializing the controller, it immediately begins interacting with its environment. At every timestep, it observes an MDP state transition tuple, calculates model residuals, uses the residuals to calculate the DRO offset $r^{(j)}(k)$, and then solves a new MPC program at the next state. This application case serves as a purposefully extreme challenge of the robustness and behavior of our algorithm at what would otherwise be unreasonable levels of uncertainty and risk. Later in our paper, we demonstrate that even under such extreme conditions, we manage to safely learn control policies for a host of nonlinear stochastic control problems. We do note, however, that our algorithm is much more widely applicable when prior data and SME is available.}
      \label{figurelabel1}
\end{figure*}

\subsection{Model Predictive Control Formulation}
We apply Wasserstein ambiguity sets to robustify a learning model predictive controller, based on the following optimization program formulation. Given true plant dynamics:
\begin{align}
x_{t+1} &= f(x_t, u_t, W_t)\\
y_{t} &= g(x_t, u_t, V_t)
\end{align}
where $t$ is the current timestep, $W_t$ is state noise, $V_t$ is output measurement noise, $x_t$ is the state variable, and $y_t$ is the output variable. We assume access to full state and output measurements, subject to the measurement noises $W_t$ and $V_t$. The capital letters represent random variables. Before considering modifications for distributional robustness to uncertainty (which also accommodate exogenous inputs), we seek to solve the following predictive control problem:
\begin{subequations}
\begin{align}
&\underset{u_{t:t+N-1}}{\text{minimize}} \sum_{k=t}^{t+N} J_k(\hat{x}_k, \hat{y}_k, u_k) \label{eqn:ftocp1}\\
&\text{subject to:}\\
&\hat{x}_{k+1} = \hat{f}(\hat{x}_k, u_k, \theta_f)\\
&\hat{y}_{k} = \hat{g}(\hat{x}_k, u_k, \theta_g)\\
&\hat{y}_k \leq 0\\
&\hat{x}_t = x_t \label{eqn:ftocp5}
\end{align}
\end{subequations}
where $x_t$ is the known (measured) initial state at the current timestep $t$. The \textit{``hat''} symbol indicates a predicted variable, and the learned models themselves are given by:
\begin{align}
    \hat{x}_{t+1} &= \hat{f}(x_t, u_t, \theta_f)\\
    \hat{y}_{t+1} &= \hat{g}(x_t, u_t, \theta_g).
\end{align}
At a high level, these can be thought of as two separate models. However, when learning a black-box representation of the system, that single model can be trained to predict both sets of values $\hat{x}_{t+1}$ and $\hat{y}_t$. The parameters $\theta_f$ and $\theta_g$ are learned from historical data through model identification.


\subsection{Model Identification}
The models are used to predict state transition dynamics and constraint function outputs. We assume the true model parameters $\theta_f^*$ and $\theta_g^*$ are inaccessible to the controller.  Several methods can be selected to learn the parameters online, and can depend on what type of learning model architecture is selected.  In this paper, we utilize nonlinear least-squares with neural network models for both the state transition dynamics and constraint functions:
\begin{align}
    \hat{f}(x_t, u_t, \theta_f) &\leftarrow x_{t+1}\\
    \hat{g}(x_t, u_t, \theta_g) &\leftarrow y_{t}
\end{align}
where $x_{k+1}$ and $y_k$ are assumed to be measurable from the real system at the current timestep. When conducting MPC, the initial $x_{k}$ is obtained by assuming full state observability throughout the LbC problem. From this point forward, we denote $\theta_{g;t}$ as the parameterization of the learned model of $g$ at timestep $t$ in the overall learning process. 


\subsection{Modeling Error Characterization}
We characterize modeling error through comprehensive modeling residuals across varying prediction depths.  

For example, consider a scalar system $x\in\mathbb{R}$, $y\in\mathbb{R}$ within three steps of model predictive control $N=2$ with quadratic, time invariant objective function (state penalty $q=1$, effort penalty $r=1$, terminal state penalty $p=1$):
\begin{subequations}
\begin{align}
&\underset{u_{t}, u_{t+1}, u_{t+2}}{\text{minimize}} \: {x}_t^2 + \hat{x}_{t+1}^2 + u_t^2 + u_{t+1}^2 + \hat{x}_{t+2}^2 \\
&\text{subject to:}\\
&\hat{x}_t = x_t\\
&\hat{x}_{t+1} = \hat{f}({x}_t, u_t, \theta_f)\\
&\hat{x}_{t+2} = \hat{f}(\hat{x}_{t+1}, u_{t+1}, \theta_f)\\
&\hat{x}_{t+3} = \hat{f}(\hat{x}_{t+2}, u_{t+2}, \theta_f)\\
&\hat{y}_{t} = \hat{g}({x}_t, u_t, \theta_g)\\
&\hat{y}_{t+1} = \hat{g}(\hat{x}_{t+1}, u_{t+1}, \theta_g)\\
&\hat{y}_{t+2} = \hat{g}(\hat{x}_{t+2}, u_{t+2}, \theta_g)\\
&\hat{y}_t \leq 0\\
&\hat{y}_{t+1} \leq 0\\
&\hat{y}_{t+2} \leq 0
\end{align}   
\end{subequations}
 
Suppose we find a sequence $u_t^*$, $u_{t+1}^*$, $u_{t+2}^*$ from solving 3 sequential model predictive control problems with the true plant in the loop. Since we are using learned models to solve these predictive control problems, these inputs are likely not actually optimal for the system, and with added PoE they include exploratory aspects.  In each case we apply the first control input to the system to obtain $x_{t+1}^*$, $x_{t+2}^*$, $x_{t+2}^*$ We can quantify prediction error of the learned constraint function in the following manner:
\begin{subequations}
\begin{align}
    R_1^{(t)} &= g(x_t, u_t^*) - \hat{g}(x_t, u_t^*, \theta_g) \\
    R_1^{(t+1)} &= g(x_{t+1}^*, u_{t+1}^*) - \hat{g}(\hat{x}_{t+1}, u_{t+1}^*, \theta_g) \\
    R_1^{(t+2)} &= g(x_{t+2}^*, u_{t+2}^*) - \hat{g}(\hat{x}_{t+2}, u_{t+2}^*, \theta_g)
\end{align}   
\end{subequations}
These are 1-step residuals, as denoted by the subscript $R_1$, since $\hat{x}_{t+1}=f(x_t, u_t^*)$ and $\hat{x}_{t+2} = f(x_{t+1}^*, u_{t+1}^*)$. In these equations, the function $g$ represents our observations from the real system (simple data), and the function $\hat{g}$ represents the predictions of our learned constraint model. We take the absolute value since these residuals will be introduced as variables that add conservatism relative to the existing constraint boundary. Since we conduct predictive control, we also want to quantify modeling errors after 2, 3, or more steps of prediction into the future using learned models, as errors can accumulate and become worse with successive prediction steps. This happens in the following way:
\begin{subequations}
\begin{align}
    R_1^{(t)} &= |g(x_t, u_t^*) - \hat{g}(x_t, u_t^*, \theta_g)| \\
    R_2^{(t)} &= |g(x_{t+1}^*, u_{t+1}^*) - \hat{g}(\hat{f}(x_t, u_{t}^*, \theta_f), u_{t+1}^*, \theta_g)| \\
    R_3^{(t)} &= |g(x_{t+2}^*, u_{t+2}^*)-\\
    &\hat{g}(\hat{f}(\hat{f}(x_t, u_t^*, \theta_f), u_{t+1}^*, \theta_f), u_{t+2}^*, \theta_g)|
\end{align}
\end{subequations}
As is shown here, modeling error accumulates from learned representation of both the constraint function $\hat{g}$ and the learned dynamics function $\hat{f}$. 

\begin{remark}
    We choose to take the absolute value of residuals. This decision is not necessary, but makes intuitive sense given the application. Since we are intending to modify the nominal constraint boundary, signals of modeling errors that show underestimation could lead to an offset that potentially moves the constraint into the unsafe region. We seek to avoid this, and only create offsets that reduce the size of the feasible region. 
\end{remark}

The model identification process utilizes the 1-step residuals to minimize mean-square prediction error (MSE) of the prediction of the state transition compared to past observations. The multi-step residuals are utilized by the DRO framework to adjust conservatism deeper into the future based on cumulative modeling error.

By representing modeling error this way, we lump all relevant sources of modeling error into an additive term. As previously discussed, the absolute value is taken as a precautionary measure. Omitting that transformation provides the following simple expression:
\begin{equation}
    g(x_{t+2}^*, u_{t+2}^*) = \hat{g}(\hat{f}(\hat{f}(x_t, u_t^*, \theta_g), u_{t+1}^*, \theta_g), u_{t+2}^*, \theta_g) + R_3^{(t)}\label{addres}
\end{equation}
By treating the residuals as random variables drawn from a true distribution $\mathbb{P}$, the constraints will by definition be additive in the random variable/modeing error.


\subsection{Safety and Robustness using Wasserstein Ambiguity Sets}
Now that we have outlined the distributionally robust chance constrained approach using the Wasserstein ambiguity set, we can describe how it fits within our robust control framework.  

The residuals defined in the previous subsection entail a representation of the modeling error. This is only true because the constraint functions are evaluated using predicted states from the learned dynamical model, whose true representation is unknown.  By considering process error/residuals as an additive noise term, we can maximize the utility of the DRO reformulation in \cite{Duan00} which requires this linear structure in the constraint:
\begin{equation}\label{eqn:ftocp3a}
     g(x_{k}, u_{k}, \theta_{g;t}) + \textbf{R}_1\leq 0
\end{equation}
As previously discussed and shown in equation (\ref{addres}), by design, this linear structure will always occur. These residuals are random variables characterized by empirical distributions based on our observations. Now, we've bolded the variable $\bf{R}_1$ to indicate it is a random variable, whereas the previous value $R_1^{(t)}$ was a realization of this random variable at time $t$.

To accommodate distributional uncertainty in our estimate of $\hat{\mathbb{P}}$, we transform the constraint (\ref{eqn:ftocp3a}) for each of $1\rightarrow N+1$ step residuals into a joint distributionally robust chance constraint via Wasserstein ambiguity set as follows:
\begin{equation}%
\begin{aligned}
     \underset{\mathbb{P} \in \mathbb{B}_\epsilon} {\text{inf}} \mathbb{P} &\left[
    \begin{array}{r}
        \hat{g}(\hat{x}_{k}, u_{k}, \theta_{g;t}) + \bf{R}_1 \leq 0 \\
        \hat{g}(\hat{x}_{k+1}, u_{k+1}, \theta_{g;t}) + \bf{R}_2 \leq 0 \\
        \vdots \\
        \hat{g}(\hat{x}_{k+N}, u_{k+N}, \theta_{g;t}) + \bf{R}_{N+1} \leq 0
    \end{array} 
    \right] \\
    & \geq 1 - \eta \label{lmpcfullcon}
\end{aligned}
\end{equation}
The reformulation we adopt from \cite{Duan00} presents a simple method to accommodate the constraint without inverting the CDF. If we operate under the assumption that the residuals for $i=1,...,N$ steps are uncorrelated, then we can decompose this joint chance constraint into a set of individual chance constraints.  This decomposition could be useful if the optimization algorithm we select to solve the MPC problem scales unfavorably with the dimension of the constraints.    Algorithm 1 provides an overview of the real-time implementation of our approach.  As previously stated, the process for computing $r$ entails a simple scalar convex optimization program.

\begin{remark}
    The reformulation from \cite{Duan00} adds cardinality of constraints that scale with order $2^m$. However, our formulation of modeling error as an additive residual allows the number of constraints to remain constant. We detail this property in the Appendix of this paper. The simple answer is that, by taking the absolute values of the residuals, the random variable that represents modeling error is strictly non-negative. This means a negative realization is impossible to encounter, and need not be accommodated. By keeping the cardinality of constraints constant, the computational scalability of our approach is preserved for higher dimensional control problems.
\end{remark}

At each time step, we compute model residuals with our most recent estimate $\theta_{g;t}$ using predicted state transitions from our entire cumulative experience, compile a unique empirical distribution $\hat{\mathbb{P}}$ corresponding to each individual chance constraint, and compute the value of $r$ in \eqref{eqn:ineq-reform} to reformulate the distributionally robust chance constraints.  We can begin the overall process with a small control horizon $N$, and gradually increase $N$ as we accumulate more and more data from experience.  The residuals we compute are for horizon lengths of $1$ to $N$-steps, meaning the elements of $\bf{R}$ correspond to each of $i = 1,...,N$ step residuals.  Then, we assemble a joint chance constraint where the elements of the column vector of the random variable are the $1\rightarrow N$ step residuals.  In \cite{Duan00}, authors pursue a DRO reformulation that utilizes a polytopic representation of the uncertainty set. Our formulation preserves scalability by isolating dependence on the random variable in the constraint. Our Appendix shows the logic that allows the cardinality of constraints to remain constant. 

Finally, when we conduct MPC, we replace the nominal constraints with their distributionally robust counterparts:
\begin{subequations}
\begin{align}
\underset{u \in \mathcal{U}}{\text{minimize}} \quad & \sum_{k=t}^{t+N} J_k(\hat{x}_k,{u}_k)  \label{eqn:drftocp1}\\
\text{s. to:} \quad
& \hat{x}_{k+1} = \hat{f}(\hat{x}_k, u_{k}, \theta_{g;t}) \\
& \begin{bmatrix}
           \hat{g}(\hat{x}_k,{u}_k, \theta_{g;t}) \\
           \hat{g}({x}_{k+1},{u}_{k+1}, \theta_{g;t}) \\
           \vdots \\
           \hat{g}(\hat{x}_{k+N},{u}_{k+N}, \theta_{g;t})
         \end{bmatrix} + r^{(j)} \leq 0 \\ 
& \hat{x}_0 = {x}_t \label{eqn:drftocp5}
\end{align}
\end{subequations}
Algorithm 1 describes the implementation of our MPC architecture coupled with the Wasserstein distributionally robust optimization scheme:
\begin{algorithm}[h]
   \caption{Wasserstein Robust Learned MPC}
   \label{alg:example}
\begin{algorithmic}
\REQUIRE  State space $\mathcal{x}$, Action space $\mathcal{U}$ \\ 


\FOR{$t$ in range $t_{max}$}

    \IF{$t = 1$}
        \item $u_t =$ known safe input, $N=1$
    \ELSE
        \item Update the dynamical system model and constraint functions $\theta_{t-1}\rightarrow\theta_t$ 
        \item Receding horizon increment rule (i.e.  $N=min\{N_{targ}, round(\frac{t}{N_{targ}})+1\}$)
        \item Obtain Wasserstein ambiguity set offset $r$:
        \item $u_t \leftarrow$ Solve MPC optimization program \eqref{eqnLbC1}-\eqref{eqnLbC5}
    \ENDIF
    \item $x_{t+1} = f(x_t, u_t, W_t)$ (Truth plant)
    \item $y_t = g(x_t, u_t, V_t)$ (Truth plant)
    \ENDFOR
\end{algorithmic}
\end{algorithm}

The MPC program specified in (\ref{eqnLbC1}-\ref{eqnLbC5}) details the slight modifications made to (\ref{eqn:drftocp1}-\ref{eqn:drftocp5}) accommodating the coupled PoE component to our LbC framework. We discuss this in more detail in part F. of this section.

One important note concerns a specific scenario of model adaptation where the true underlying system slowly changes.  Our application of receding horizon control necessitates the use of a snapshot model in the prediction phase.  This requires we assume the rate of change of the dynamics of the true plant is relatively small. In such conditions, however, the historical residuals we collect through measurements will slowly lose relevance.  This issue can be easily reconciled with use of either a moving window of residuals, or with a proper forgetting scheme.  In this paper, we propose a simple method to accommodate such cases. Since the focus of this paper is on \textit{tabula-rasa} learning-based control, we relegate the discussion of this additional framework to this paper's appendix.

\subsection{Horizon Increment Rule}

MPC with well-defined dynamical structure can leverage judicious selection of the prediction horizon as a component to proving recursive feasibility. When considering a general class of systems as is the case with MBRL, the prediction horizon becomes a hyperparameter that manages the tradeoff between prediction depth and computational expense.  In this paper, we elect to define a simple horizon increment rule for our experiments. Typically in learning-based control, the prediction horizon is a hyperparameter whose selection can be done empirically with more nuanced methods \cite{bandmbrl, hyperband}. In our case studies, which we design to emulate \textit{tabula-rasa} learning-based control as closely as is consistent with the assumptions of our algorithm, we utilize this horizon increment rule as a heuristic to simply allow the problem to be rapidly solved. By solving severely restrictive case studies, we validate the performance of our method under the most challenging context for which it is technically designed. For real-world applications, the horizon can often be selected using a combination of available subject matter expertise (which should not be ignored if it is available), and automatic tuning methods like those of \cite{bandmbrl, hyperband}. The increment rule is not meant as a serious method for real-world embedded control systems that often possess highly limited computational resources.



\subsection{Persistence of Excitation, and Problem Assumptions } 

This subsection defines the set of least restrictive assumptions we identify towards achieving safe learning-based control. In this paper, we consider systems with non-hybrid dynamics for simplicity. Our method leverages proved safety properties from \cite{Duan00}, which apply to static optimization programs. We identify that these methods can apply to LbC problems under a series of assumptions made in this section. These assumptions almost entirely relate directly to situations when the dynamical, DRO, and PoE components, which are normally not considerations for static optimization programs, could create opportunities for empty feasible sets. This subsection defines a PoE scheme directly amenable to translating guarantees from \cite{Duan00} to our formulation.  Notably, our assumptions are significantly less restrictive than those of existing LbC methods.   The majority of these assumptions relate to clear necessary conditions which we detail here:

\begin{assumption}
A feasible state and control trajectory exists for each prediction horizon $N$ in the optimal control problem.
\end{assumption}
This is the most fundamental requirement to apply safe control. 

\begin{assumption}
We assume we know a safe control input which we can apply at the first timestep.
\end{assumption}
Starting with limited model knowledge, if we don't know a temporarily safe control input we can apply at the first timestep, we obviously can't translate any meaningful safety certificates. This contrasts to other work which requires knowledge of safe control trajectories throughout the time horizon, or a known safe backup policy.

\begin{assumption}
Starting with an optimal control problem of the form (\ref{eqn:ftocp1}-\ref{eqn:ftocp5}), suppose we have a constraint function $g(x_{k}, u_{k}, \theta_{g;t}) : \mathcal{x} \times \mathcal{U} \times \theta \rightarrow \mathcal{S}$.  The sublevel set $\mathcal{G}_{r_{DRO}} = \{ (x,u) \in \mathcal{x},  \mathcal{U} \: : \: g(x,u) + r_{DRO} \leq 0\}$ defines the adjusted feasible region, where feasibility is satisfied at the current timestep. This set must not be empty $\forall r_{DRO} \in \mathcal{R}$, where the set $\mathcal{R} = \{ r_{DRO} \in \mathbb{R} \: : \: 0 \leq r_{DRO} \leq r_{DRO;max} \} $ describes the set of all potential values of the DRO offset.
\end{assumption}
Since our method relies on creating an offset from the nominal constraint boundary, any potential value of the offset must lie in the image of the constraint function. 

This assumption can be thought of as a generalization of a common LbC assumption that relates to ``bounded modeling error,'' an example of which is given by Assumption 2 in \cite{Berk00}.  In our case, using general function approximation, our method to quantify model error is empirically based on residuals.  If the residuals of the learned model are too large, indicating our learned model is inaccurate, the resulting computed $r_{DRO}$ (which is a conservative approximation of the residual, based on its distribution) will enforce a large offset from the nominal boundary. This assumption says that if the learned model is sufficiently inaccurate, the offset will be so large that the adjusted feasible region is empty, which is incompatible with the setup of \cite{Duan00}. The value $r_{DRO;max}$ represents any maximum residual value we can potentially infer from the problem, and can be defaulted to as an empirical approach if this case is reached in a real problem, although safety properties may not be reliable in such cases. Our experiments show such scenarios can be unlikely to occur, although the possibility of their occurrence should be considered.




The next assumption relates to a slightly stronger condition regarding persistence of excitation (PoE). The agent must be capable of exploring during LbC. In order to ensure the guarantees from \cite{Duan00} translate under those diverse circumstances, the same statements of 3.1-3.3 must be satisfied with respect to an additional exploration process $\mathcal{N}$ that ensures PoE.  

For clarity, we define the following modified MPC program that considers an additive exploration signal from $\mathcal{N}$:

\begin{subequations}
\begin{align}
\underset{{u}, u^n \in \mathcal{U}}{\text{minimize}} \quad & \sum_{k=t}^{t+N} J_k(\hat{x}_k,{u}_k)  \label{eqnLbC1}\\
\text{s. to:} \quad
& \hat{x}_{k+1} = \hat{f}(\hat{x}_{k}, u_{k}, \theta_{g;t}) \\
& \hat{x}^n_{k+1} = \hat{f}(\hat{x}^n_k, u^n_k, \theta_{g;t}) \\
& \hat{g}(\hat{x}_{k}, u_{k}, \theta_{g;t}) + r_{DRO}\leq 0\\
& \hat{g}(\hat{x}^n_k, u^n_k, \theta_{g;t}) + r_{DRO}\leq 0\\
& u^n = {u} + N_{i:i+N}\\
& N_{i:i+N} \sim \mathcal{N}\\
& \hat{x}_0 = {x}_t\\
& \hat{x}^n_0 = {x}_t\label{eqnLbC5}
\end{align}
\end{subequations}
where $\mathcal{N}$ is the distribution of a random exploration process which can be added to the nominal control input, and the superscript $x^n$ and $u^n$ denote trajectories perturbed by the exploration signal. The solution $u^n(t)^\star$ is then applied to the plant at time step $t$. 
\begin{remark}
Equations (\ref{eqnLbC1}-\ref{eqnLbC5}) guarantee feasibility from $k=t$ to $k=t+N$ for a system with parameters $\theta_{g;t}$ with a specified risk metric/probabilistic guarantee. This is formulated to guarantee feasibility over the control horizon. To assess recursive feasibility, one could utilize the methods from \cite{MPC_new00,deepc00} that require more significant restrictions in the form of model knowledge, mathematical structure on the feedback policy, and prior existing safe data. 
\end{remark}

The additive noise perturbation for exploration takes inspiration from common methods with actor-critic or policy gradient learning, where noise via an Ornstein-Uhlenbeck process is added to the control input \cite{lillicrap2015continuous}. Relative to those existing methods, we make the following modifications for implementation:
\begin{remark}
We must constrain both nominal and perturbed trajectories to ensure safety even with exploration.  If we only add the perturbation after solving the MPC program, safety is not guaranteed. 
\end{remark}
\begin{remark}
A scalarized tradeoff between $J_k(\hat{x}_{k},u_{k})$ and $J_k(\hat{x}^n_k,u^n_k)$ can be formulated to balance exploration and exploitation during planning.
\end{remark}

Now, we define the next assumption relevant to translating safety to LbC systems under strong limitations on SME:
\begin{assumption}
Given the noise process $\mathcal{N}$ defined to satisfy PoE for the model identification problem, the constraints $g(x_{k}, u_{k}, \theta_{g;t})$ and $g(x^n_k, u^n_k, \theta_{g;t})$ of the snapshot model must be satisfied for every realization from $\mathcal{N}$ throughout the overall finite-time optimal control problem.
\end{assumption}

Given these conditions, we state the following remark detailing the properties of our method:
\begin{remark}
Based on the provided safety guarantee afforded from the adopted DRO framework from \cite{Duan00}, (\ref{eqn:drftocp1}-\ref{eqn:drftocp5}) admits a feasible solution that satisfies the nominal constraints w.p. $1-\eta$ as long as the feasible set is not empty, which follows from Assumptions 3.1-3.4. 
\end{remark}

We also state two remarks that help with implementation of our approach.
\begin{remark}
These assumptions must also hold for the prediction horizons chosen at each instant in time.
\end{remark}
\begin{remark}
If the DRO offset is so large it creates an empty feasible set, an artificial value $r_{DRO;max}$ can be defaulted to to facilitate implementation, although safety guarantees in such situations may be difficult to translate. If a random search is used to solve the MPC program in such cases, the evaluated trajectory that creates the least predicted constraint violation given the unmodified DRO offset can be selected.
\end{remark}

\section{Case Study in Safe Online Lithium-Ion Battery Fast Charging}
In this section, we validate our approach using a nonlinear lithium-ion battery fast charging problem. This problem closely emulates the performance-safety tradeoffs of common safe RL validation studies including ant-circle \cite{Achiam00}. Specifically, the objective is to charge the battery cell as fast as possible, but the charging is limited by nonlinear voltage dynamics which must stay below critical thresholds.  Violation of the voltage constraint can lead to rapid aging and potential catastrophic failure. However, higher input currents (which increase voltage) also directly charge the battery more rapidly.  Thus, the optimal solution is a boundary solution where the terminal voltage rides the constraint boundary.  This presents a problem with significant challenges and tradeoffs relating to safety and performance.  Exploring how such algorithms accommodate these challenges can reveal insights into their overall efficacy and shortcomings.

\begin{table*}[t]
\caption{UPDATE VALUES TO BE CONSISTENT WITH REPO CODE Run Safety, computational, and performance comparison for DRO-MPC and MPC with battery fast charging. Activation of the DRO offset begins at \tt\small minResidNum = 2.}
\label{batt_dat_ini}
\begin{center}
\begin{small}
\begin{sc}
\begin{tabular}{ccccc}
\hline \hline
(DRO) & \% Violations [\%] & Max Voltage [V] & Iteration Time [s] & Charging Time [min] \\ 
\hline
1 & 0.0 \% & 3.5944 & 0.8551 & 7.3833  \\
2 & 0.4 \% & 3.7004 & 0.8473 & 7.7667 \\
3 & 0.2 \% & 3.6887 & 0.8529 & 7.3000 \\
4 & 0.6 \% & 3.7098 & 0.8503 & 8.1833 \\
5 & 0.0 \% & 3.5927 & 0.8688 & 7.5333 \\
6 & 0.4 \% & 3.7344 & 0.8550 & 7.7833 \\
7 & 0.4 \% & 3.7032 & 0.8643 & 8.1167 \\
8 & 0.2 \% & 3.6921 & 0.8692 & 7.6667 \\
9 & 0.2 \% & 3.6916  & 0.8620 & 7.8667 \\
10 & 0.2 \% & 3.6985 & 0.8375 & 8.0167\\
\hline
Averages &0.26\% & 3.6806 & 0.8562 & 7.8150 \\
\hline \hline
Run (no DRO) & \% Violations [\%] & Max Voltage [V] & Iteration Time [s] & Charging Time [min] \\
\hline
1 & 4.2 \% & 3.7795 & 0.8630 & 6.8667 \\
2 & 7.4 \% & 3.7604 & 0.8345 & 6.8667 \\
3 & 5.0 \% & 3.7474 & 0.8055 & 6.7833 \\
4 & 13.6 \% & 3.7284 & 0.7938 & 6.8500  \\
5 & 8.0 \%  & 3.9072 & 0.8020 & 6.8333 \\
6 & 16.2\% & 3.9060 & 0.7977 & 6.8667  \\
7 & 8.0 \% & 3.9040  & 0.8240 & 6.8667  \\
8 & 11.6 \% & 3.7651 & 0.7875 &  7.0167 \\
9 & 7.2 \% & 3.7736 & 0.8237 & 6.8000  \\
10 & 16.4 \% & 3.7634 & 0.7928 & 6.7500  \\
\hline
Averages & 9.76 \% & 3.8035 & 0.8125 & 6.8500 \\
\hline \hline
\end{tabular}
\end{sc}
\end{small}
\end{center}
\vskip -0.1in
\end{table*}
\subsection{Equivalent Circuit Model of a Lithium-Ion Battery}
Lithium-ion batteries can be modeled with varying degrees of complexity.  Some of the more detailed dynamical models are based on electrochemistry.  For example, the Doyle-Fuller-Newman (DFN) electrochemical battery model is a high-fidelity first-principles derived physics based model of the dynamics within a lithium-ion battery \cite{Doyle00}.  Varying model-order reduction can be applied, yielding versions including the single particle model and the equivalent circuit model (ECM).  For simplicity, this paper's case study utilizes an ECM. The relevant state variables in this model are the state of charge $SOC$ and capacitor voltages $V_{RC}$ in each of two RC pairs. The relevant constraint is on the terminal voltage $V$.  This constraint prevents the battery from overheating or aging rapidly during charging and discharging. The state evolution laws are given by: 
\begin{align}
    SOC_{k+1} &= SOC_k + \frac{1}{Q}I_k\cdot \Delta t \label{eqn:1a} \\
    V_{\text{RC}_1;k+1} &= V_{\text{RC}_1;k} - \frac{\Delta t}{R_1 C_1}V_{\text{RC}_1;k} + \frac{\Delta t}{C_1}I_k \\
V_{\text{RC}_2;k+1} &= V_{\text{RC}_2;k} - \frac{\Delta t}{R_2 C_2}V_{\text{RC}_2;k} + \frac{\Delta t}{C_2}I_k \\
V_k=V_{\text{ocv}}(&SOC_k) + V_{\text{RC}_1;k} + V_{\text{RC}_2;k} + I_k R_0  \label{eqn:2a}
\end{align}
where $I(t)$ is the current input (which is the control variable for this problem), and $V_{OCV}$ is the open-circuit voltage function, which is conventionally measured through experiments.  The full experimental OCV curve is used to represent the true plant in the loop, and is obtained from a lithium-iron phosphate (LFP) battery cell  \cite{Perez05}.  In this paper, we learn the dynamics of the states and output using a simple feed-forward neural network model.




\begin{table}[t]
\caption{Relevant Parameters}
\label{sample-table2}
\begin{center}
\begin{small}
\begin{sc}
\begin{tabular}{lcccr}
\hline \hline
Parameter & Description & Value & Units \\
\hline
$Q$    & Charge Capacity & 8280 & $[\frac{1}{A.h}]$ \\
$R_0$    & Resistance & 0.01 &  $[\Omega]$      \\
$R_1$ & Resistance & 0.01 & $[\Omega]$\\
$R_2$    & Resistance & 0.02 &  $[\Omega]$      \\
$C_1$    & Capacitance & 2500 & $[F]$ \\
$C_2$    & Capacitance & 70000 & $[F]$ \\
$\Delta t$ & Timestep & 1 & [s] \\
$N_{targ}$ & Max Control Horizon & 8 & [-] \\
$\eta$ & Risk Metric & 0.025 & [-] \\
$\beta$ & Ambiguity Metric & 0.99 & [-] \\
$SOC_0$ & Initial SOC & 0.2 &[-] \\
$SOC_{targ}$ & Target SOC & 0.8 & [-] \\
$V_{RC_1}(0)$ & Init. Cap. 1 Voltage & 0 & [V] \\
$V_{RC_2}(0)$ & Init. Cap. 2 Voltage & 0 & [V] \\

\hline \hline
\end{tabular}
\end{sc}
\end{small}
\end{center}
\vskip -0.1in
\end{table}

\subsection{Model-Predictive Control Formulation}

We utilize the following formulation of fast charging:
\begin{equation}\label{eqn::mpc}
\underset{I_k \in \mathcal{U}}{\text{minimize}}  \sum_{k=t}^{t+N} (SOC_k - SOC_{target})^2  
\end{equation}
subject to:
\begin{align}
(\ref{eqn:1a})-(\ref{eqn:2a}), &\quad
SOC(0) = SOC_0 \\
V_k \leq 3.6 V, &\quad
0 A \leq I_k \leq 40 A\label{eqn::mpc2}
\end{align}


\begin{remark}
In our case, we assume the controller does not have access to the form of the underlying dynamics given by (\ref{eqn:1a}-\ref{eqn:2a}). Instead, we apply our end-to-end LbC method to learn the dynamics ``from scratch'' as is consistent with \textit{tabula-rasa} learning methods. We utilize neural network black-box models to accomplish this. The rules used to update the neural network parameters affect the convergence of the data-driven model to accurate behavior, which also effects empirical safety. We keep the neural network training consistent between our DRO algorithm and its non-robust baseline. The exact training procedure can be referenced in the public codebase \cite{KandelGitHub}. Updating more slowly at first tends to encourage more safe behavior.
\end{remark}

In these case studies, we apply perturbation to the inputs that further excite the system, towards ensuring PoE. These perturbations are drawn as uniform vectors whose elements lie between $-2.5 \leq x_p \leq 2.5$ Amps. These perturbations are applied to both the distributionally robust controller, as well as the non-robust baseline controller In both cases, we seek to ensure mutual constraint satisfaction for the trajectories predicted using both the nominal and perturbed inputs.

We only allow a maximum total of 500 seconds for the battery to be charged.  The timestep $\Delta t=1$ seconds, $\eta=0.025$, $\beta = 0.99$, and $N_{targ}=8$ steps. Our neural network dynamical model has 1 hidden layer with 3 neurons and sigmoid activation function, with a linear output layer. To solve the MPC problem, we apply a $(1+\lambda)$ evolutionary strategy (ES) based on a normally distributed mutation vector. In our appendix, we describe how this strategy works, why we select it, and other reasonable alternatives. The solver works with a single iteration and 250,000 mutants.  The initial point of the ES is taken as the optimal point from the previous timestep. Addressing Assumption 2, we assume that at the first timestep, control inputs of $I_k \leq 25$ Amps are known to be temporarily safe. Since we constrain voltage which is a scalar, the constraint function dimension $m=1$. 

Our baseline is a learning MPC controller with no DRO framework.  We adopt the same problem formulation as if we were going to add the constant $r_{DRO}$ to the constraints, but we omit the DRO constant in the end to evaluate the impact it has on the robustness of the final control law.

\subsection{Results}
\begin{figure*}[ht!]\label{fig:res7}
      \centering  
      \includegraphics[trim = 0mm 0mm 0mm 0mm, clip, width=\textwidth]{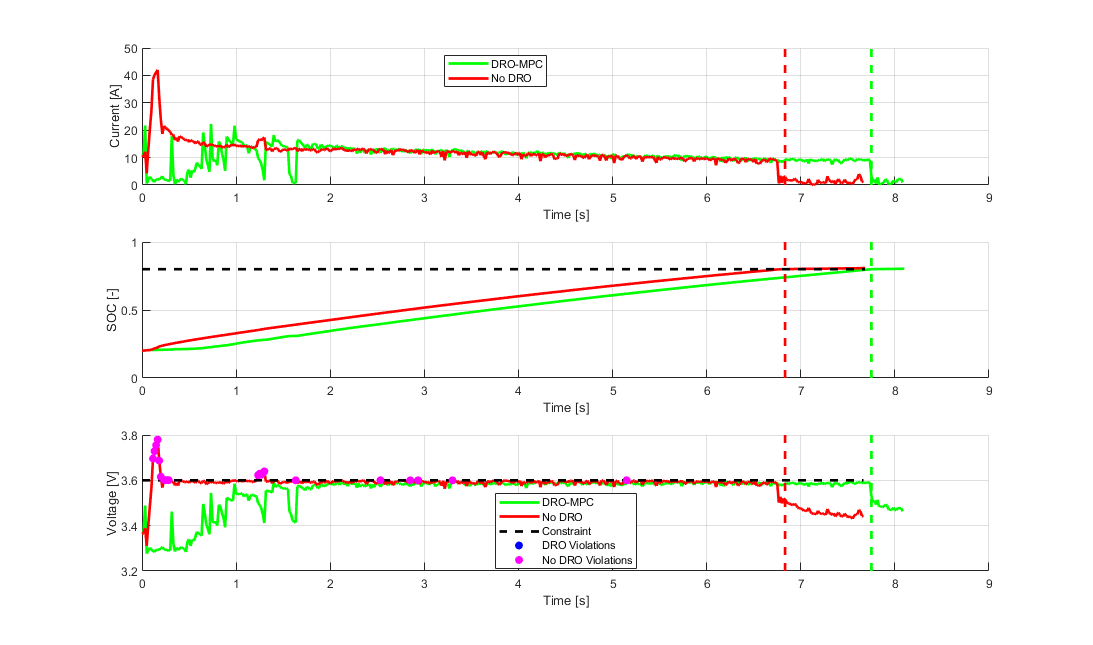}\label{run1}
      \caption{Comparison of nonlinear MPC Controller with and without DRO for lithium-ion battery fast charging. Run 1 is shown here. }
      \label{figurelabel3}
\end{figure*}
\begin{figure*}[ht!]\label{fig:res2}
      \centering  
      \includegraphics[trim = 0mm 0mm 0mm 0mm, clip, width=\textwidth]{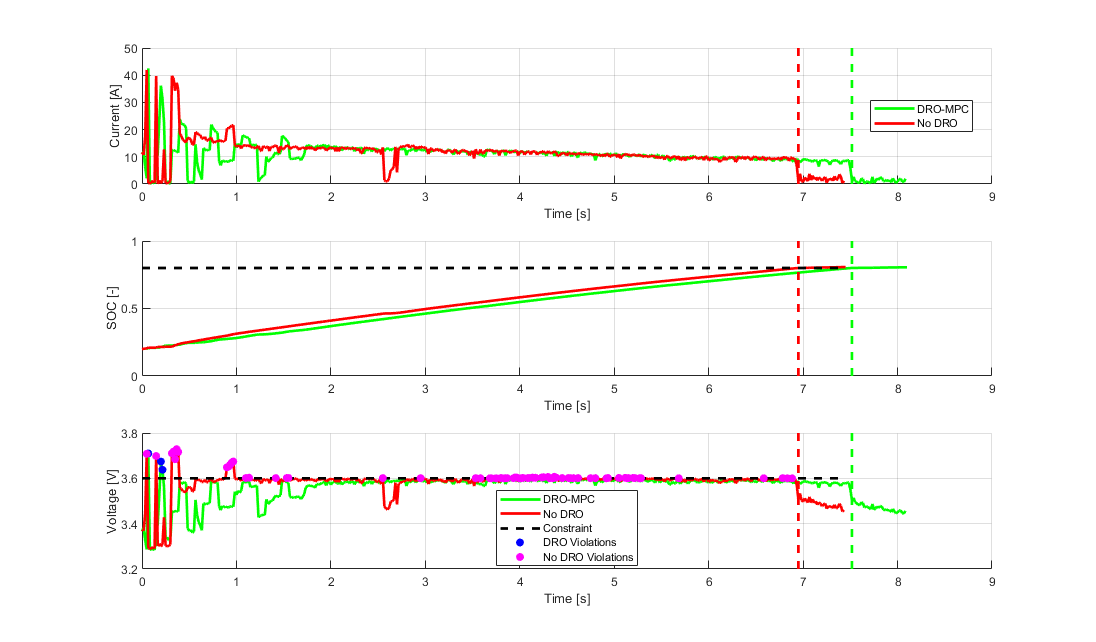}
      \caption{Comparison of nonlinear MPC Controller with and without DRO for lithium-ion battery fast charging. Run 4 is shown here.}
      \label{figurelabel4}
\end{figure*}
In total, we conducted a series of 10 experiments with identical designs but different initial random seeds. We run our algorithm and a non robust baseline for these 10 independent runs on the same battery fast charging problem detailed in the previous subsections.  Table 1 shows the performance, computation, and safety statistics for each of these runs.  For a closer look, we go to Figure 2 which shows one run of both the DRO algorithm and its non-robust counterpart.  In the case of Figure 2 (run 1), the DRO-based does not violate the constraint at any point.  In Figure 3 we see the highest incidence of constraint violation for the DRO controller (from run 4).    Conversely, the non-robust versions both experiences a combination of initial, significant voltage spikes as well as minor violations which persist throughout the experiments.  In total, if we focus on Figure 3 (run 4), the non-robust version violated constraints in 13.6 \% of timesteps (68 timesteps out of 500 total).  The charging time was 6.85 minutes, which was 16.29\% faster than the DRO version, whose charging time was 8.1833 minutes.  This makes intuitive sense, as the added DRO framework introduces additional conservatism which affects the performance of the overall control policy.  

Overall across all 10 runs, our DRO version violates constraints in 0.26\% of total timesteps, which is well within the chosen value of $\eta = 0.025 = 2.5\%$ over just a single optimization iteration.  The non-robust version, however, violates constraints in 9.76\% of total timesteps on average. Similarly, there is a stark difference in the maximum voltages seen by the robust and non-robust versions, with the DRO framework reducing the mean peak voltage by 122.9 millivolts.  The DRO calculations increase the overall computation time by an average of 43.7 milliseconds per timestep, and allow the algorithm in this case to run in real time.  No optimizations were made to the Matlab code to expedite the runtime of either algorithm, and the only difference in code between the two algorithms is the auxiliary and separate DRO framework.  Finally, across the 10 total runs the overall charging time with the DRO framework averages 7.8150 minutes, approximately 14.1\% longer than that of the non-DRO version. Given the safety-critical nature of this control problem, the safety guarantees of our algorithm are likely well worth the marginal degradation to the charging performance resulting from added conservatism.


\section{Case Study in Safe Autonomous Driving and Obstacle Avoidance}

In the following section, we implement our algorithmic architecture to safely learn to drive a vehicle while avoiding obstacles.  This learning occurs within the same design as our battery case study, namely we begin with zero model knowledge and only a single known safe control input.  We fit a data-driven model to the dynamics and conduct receding-horizon control.  

This study is designed with specific decisions in mind to more effectively reveal the efficacy of our algorithm.  Some of these decisions make our study somewhat unrealistic insofar as they expose the agent to greater danger than necessary.   The following subsections discusses these decisions in more detail.

\subsection{Dynamical Simulator}
In this case study, we utilize a bicycle model for the vehicle dynamics.  This environment is encoded in the following equations discretized via forward Euler approximation:

\begin{align}
    x_{1;t+1} &= x_{1;t} + \Delta t(x_{4;t} \cos(x_{3;t}) )\label{byc2::1}\\
    x_{2;t+1} &= x_{2;t} + \Delta t(x_{4;t}\sin(x_{3;t}))\\
    x_{3;t+1} &= x_{3;t} + \Delta t\left(x_{4;t} \frac{\tan(u_{2;t})}{L}\right)\\
    x_{4;t+1} &= x_{4;t} + \Delta t(u_{1;t}).\label{byc2::4}
\end{align}
where $t$ is the current timestep, $x_1$ and $x_2$ are the x-y position of the vehicle, $x_3$ is the vehicle heading angle, $x_4$ is the vehicle velocity, $u_1$ is the acceleration input (in $\frac{m}{s^2}$), and $u_2$ is the steering angle input in radians. These equations represent the true plant, which is unknown to our learning-based controller.

\subsection{Model Predictive Control Formulation}
We utilize the following formulation of simple autonomous driving with obstacle avoidance:
\begin{equation}
\underset{u_{k} \in \mathcal{U}}{\text{minimize}} -(x_1(t+N) + x_2(t+N)) \label{obj:form}
\end{equation}
subject to:
\begin{align}
(\ref{byc2::1})-(\ref{byc2::4}), &\quad x(0) = x(t) \\
Z(x_{k}) \leq Z_{cutoff}, &\quad
u_{min} \leq u_{k} \leq u_{max}\label{eqn::avmpc2}
\end{align}

Here, $Z(x_{k})$ is the obstacle barrier function which we limit to be smaller than a specified value (corresponding to the definition of the edge of the obstacle).  Residuals in the DRO algorithm are with respect to this barrier function using predicted values of the dynamical state, as opposed to the value of the obstacle function obtained with the true state. We create the driving environment defined by $Z(x_{k})$ by generating and summing random Gaussians in 2 dimensions.  Then, we define the obstacle boundaries by setting a threshold within the static map, below which becomes the safe region and above which the obstacles inhabit.  This map is used with interpolation during the final experiment.  If this constraint is violated, the agent will take actions which minimize constraint violation until feasibility is restored.  We set $u_{min} = [-1, -0.75]$, $u_{max} = -u_{min}$.  The experiment terminates once the vehicle leaves the 100 $\times$ 100 meter space. 

With the learned neural network dynamics models, the MPC formulation in (\ref{obj:form}-\ref{eqn::avmpc2}) becomes:
\begin{equation}\label{eqn::avxmpc2}
\underset{u_{k} \in \mathcal{U}}{\text{minimize}} -(\hat{x}_1(t+N) + \hat{x}_2(t+N))
\end{equation}
subject to:
\begin{align}
\hat{x}_{k+1} &= f^{NN}(x_{k}, u_{k}, \theta)\\
\hat{x}(0) &= x(t) \\
Z(\hat{x}_k) &\leq Z_{cutoff} - r_{DRO} \\
u_{min} &\leq u_{k} \leq u_{max}\label{eqn::avxmpc3}
\end{align}
Table 3 includes relevant parameters of our case study design. In this case study, we simply use 1-step residuals by relying on a basic assumption that the modeling error is uncorrelated to the depth of prediction.  Based on our experiments, this assumption is reasonable.  

 \begin{table}[t]
\caption{Relevant Parameters}
\label{sample-table3}
\begin{center}
\begin{small}
\begin{sc}
\begin{tabular}{lcccr}
\hline \hline
Parameter & Description & Value & Units \\
\hline
$L$ & Vehicle Length & 0.5 & [m] \\
$\Delta t$ & Timestep & 0.2 & [s] \\
$N_{targ}$ & Max Control Horizon & 12 & [-] \\
$\eta$ & Risk Metric & 0.005 & [-] \\
$\beta$ & Ambiguity Metric & 0.99 & [-] \\
$x_1(0)$ & Initial x-position & 5 &[m] \\
$x_2(0)$ & Initial Y-position & 10 &[m] \\
$x_3(0)$ & Initial vehicle angle & $\frac{\pi}{4}$ &[rad] \\
$x_4(0)$ & Initial velocity & 0.5 &[m/s] \\
\hline \hline
\end{tabular}
\end{sc}
\end{small}
\end{center}
\vskip -0.1in
\end{table}

\begin{table*}[t]
\caption{Safety comparison for DRO-MPC and MPC with vehicle obstacle avoidance. The max violation is in terms of the Euclidean distance. The numbers in parenthesis are the total number of timesteps where constraints are violated, with the denominator being the number of timesteps before the vehicle leaves the 100 $\times$ 100 sized environment.}
\label{sample-table4}
\begin{center}
\begin{small}
\begin{sc}
\begin{tabular}{ccccc}
\hline \hline
Run & \% Violations (DRO) & Max Violation (DRO) [m]  &\% Violations (no DRO) & Max Violation (no DRO) [m]  \\ 
\hline
1 & 0\% (0/156) & 0 & 2.05 \% (3/146) & 0.3877 \\
2 & 0 \% (0/145)& 0 & 0.65 \% (1/155) & 0.0121 \\
3 & 0.57\% (1/174)  & 0.0386  & 3.47 \% (5/144) & 0.4472 \\
4 & 0 \% (0/184)& 0  & 7.94 \% (17/214) & 0.9986 \\
5 & 0 \% (0/167)& 0  & 1.12 \% (2/179)  & 0.1897 \\
6 & 0 \% (0/140) & 0 & 8.55 \% (23/269) &  2.6259 \\
7 & 0 \% (0/148)& 0  & 6.74 \% (13/193) & 1.6726\\
8 & 0 \% (0/143)& 0 & 4.73 \% (8/169) & 0.2581 \\
9 & 0 \% (0/182)& 0  & 10.27 \% (23/224) &  1.1720 \\
10 & 0 \%  (0/165)& 0 & 1.14 \% (2/175) & 0.1772\\
\hline
Averages &0.0623\% & 0.00386 & 5.193 \% & 0.8041  \\
\hline \hline
\end{tabular}
\end{sc}
\end{small}
\end{center}
\vskip -0.1in
\end{table*}

\begin{figure*}[ht!]\label{fig:res3}
      \centering  
      \includegraphics[trim = 0mm 0mm 0mm 0mm, clip, width=\textwidth]{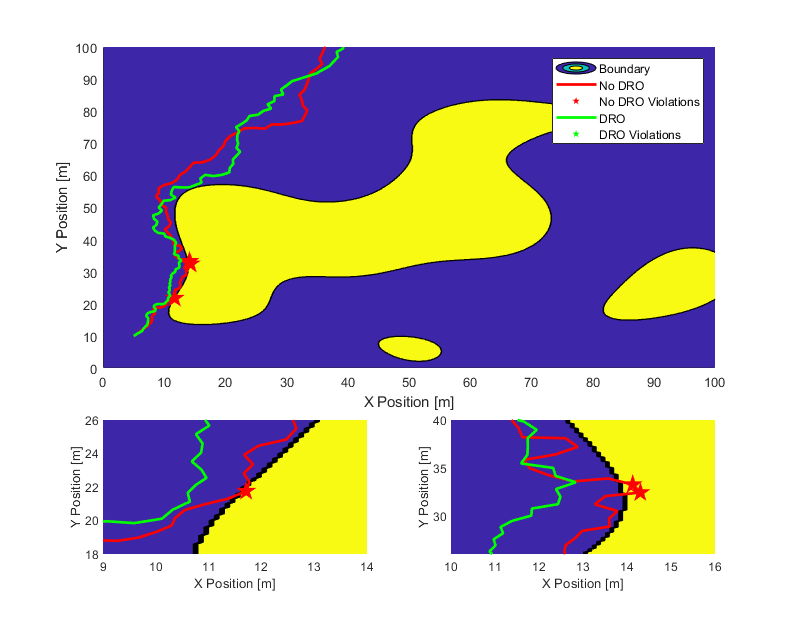}
      \caption{Comparison of nonlinear MPC Controller with and without DRO for vehicle obstacle avoidance. In this run, the DRO controller does not violate the constraints at all. This figure shows run 1, with the bottom plots revealing close ups of the areas with the highest constraint violation.}
      \label{figurelabel5}
\end{figure*}

\begin{figure*}[ht!]\label{fig:res4}
      \centering  
      \includegraphics[trim = 0mm 0mm 0mm 0mm, clip, width=\textwidth]{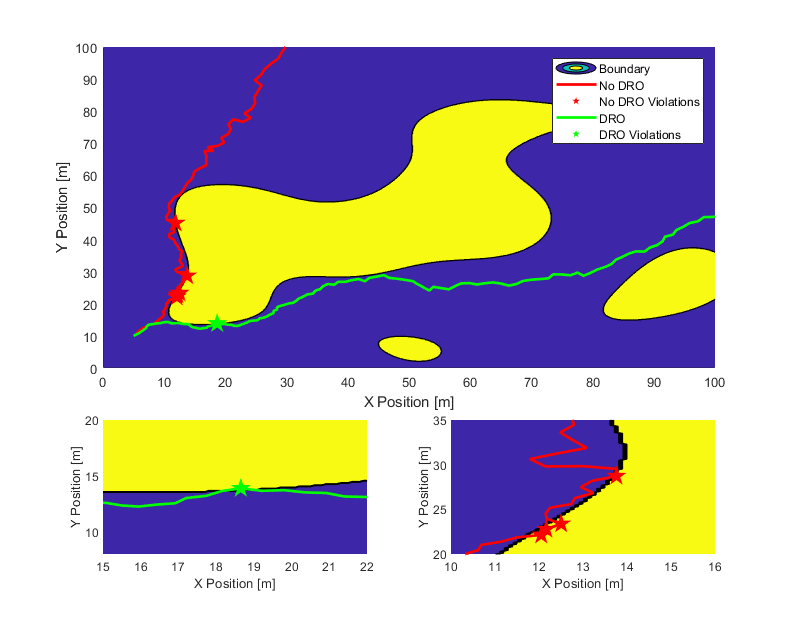}
      \caption{Comparison of nonlinear MPC Controller with and without DRO for vehicle obstacle avoidance. This figure shows run 3, with the bottom plots revealing close ups of the areas with the highest constraint violation.}
      \label{figurelabel6}
\end{figure*}

We make a deliberate choice for this objective function for a host of reasons.  While it necessarily encodes our intended behavior, it also is simple and at odds with the preeminent objective of avoiding obstacles.  Normally, we might want to encode additional considerations to constraints.  However, by allowing our simple objective function to drive the vehicle directly towards the obstacles, our control algorithm must be capable of managing the vehicle while simultaneously maintaining safety throughout most of the experiment.  Thus, this case study is designed to specifically focus on the added safety contributions from the DRO framework.

For our learned model, we initialize a feed forward neural network based on a single hidden layer with 10 neurons.  The hidden layer uses sigmoid activation functions, and the output layer uses linear activation. At the first timestep, we assume control inputs of a zero vector are known to be safe.  To solve the MPC problem, we use the same $(1+\lambda)$ evolutionary strategy used in our battery case study.  In this case, we modify the optimization algorithm such that we utilize 750,000 mutants.  We also increase the maximum prediction horizon to $N_{max}=12$ to improve the consistency of our results.

\subsection{Results}

Much like our battery case study, we conduct 10 individual runs with both our algorithm and a non-robust version.  Figures 4 and 5 show runs 1 and 3, respectively.  Table 4 shows the safety statistics from the total set of experiments.

We observe marked improvements to safety with use of our DRO algorithm.  With the DRO controller, only 1 of the 10 total runs violates constraints at all and only during a single timestep. The overall violation with the DRO controller is 0.0623\% of timesteps. Moreover, the magnitude of the violation with the DRO controller is equivalent to the vehicle skimming the edge of the boundary by less than 0.0386 meters.  Conversely, the non robust controller shows significant constraint violation in nearly all 10 runs.  The constraint violation of the non robust controller averages 0.8041 meters of violation, which represents a complete collision with the obstacle (given our vehicle length $L=0.5$).  Furthermore in one run, the non robust controller drives the vehicle nearly 3 meters into the boundary before correcting and exiting the unsafe region. 

To verify the model is operating in nonlinear portions of the state space, Figure 6 shows the range of the variable $x_3$ throughout experiment 1.
\begin{figure}[ht!]\label{fig:res5}
      \centering  
      \includegraphics[scale=0.45]{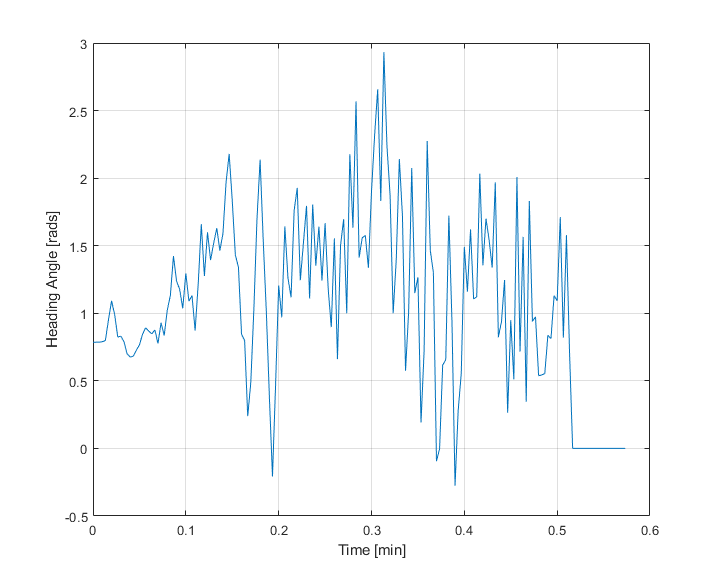}
      \caption{Heading angle trajectory for run 1 (same as that shown in Figure 6). The total range of heading angles is nearly $\pi$, showing exploration of highly nonlinear portions of the state space. The feasible range of steering angle input also covers a range of nonlinear behavior in the dynamics.}
      \label{figurelabel7}
\end{figure}


\section{Discussion}
In these case studies, we have not only explored the behavior of our algorithm at the boundary of available knowledge and data, but have validated the theoretical safety properties of our approach under the most challenging arena of its applicability.  Importantly, our approach is widely relevant in many LbC contexts. For real-world applications, we are unlikely to conduct this restrictive type of \textit{tabula-rasa} LbC. However, the same safety guarantees we have rigorously validated in these case studies are similarly applicable when more data and knowledge is available (e.g. conventional adaptive control, but with the modeling capacity of nonlinear machine-learning models).  

Since our approach functions as an end-to-end LbC method, it is amenable to more unconventional applications including control synthesis from images or any sort of state embedding \cite{visuo}.  Since we leverage black-box modeling to predict state transitions, as long as we can formulate constraints from the available state representation, we can apply our method for LbC with probabilistic safety guarantees. We relegate exploration of our method for embedding-based LbC to future work.

\section{Conclusion}
This paper presents an end-to-end distributionally robust model-based control algorithm. It addresses the problem of safety during learning-based control with strong limitations on our available knowledge and subject matter expertise.  We adopt a stochastic MPC formulation where we augment constraints with random variables corresponding to empirical distributions of modeling residuals. By applying Wasserstein ambiguity sets to optimize over the worst-case modeling error, we translate an out-of-sample safety guarantee subject to new data and experience. We validate this finding through simulation experiments.  This method is applicable to nonlinear MPC, but when applying to convex MPC programs it preserves convexity of the optimization program. 

Our results provide the basis for several meaningful insights.  It is clear that the supporting research for Wasserstein ambiguity sets provide an ideal base for its application to learning-based control.   Our numerical experiments indicate our approach is highly effective at providing probabilistic safety guarantees even in challenging cases of online learning-based control nearly from scratch. 






%
\bibliographystyle{./IEEEtran} 
\bibliography{./root}

\begin{thebibliography}{10}
\providecommand{\url}[1]{#1}
\csname url@rmstyle\endcsname
\providecommand{\newblock}{\relax}
\providecommand{\bibinfo}[2]{#2}
\providecommand\BIBentrySTDinterwordspacing{\spaceskip=0pt\relax}
\providecommand\BIBentryALTinterwordstretchfactor{4}
\providecommand\BIBentryALTinterwordspacing{\spaceskip=\fontdimen2\font plus
\BIBentryALTinterwordstretchfactor\fontdimen3\font minus
  \fontdimen4\font\relax}
\providecommand\BIBforeignlanguage[2]{{%
\expandafter\ifx\csname l@#1\endcsname\relax
\typeout{** WARNING: IEEEtran.bst: No hyphenation pattern has been}%
\typeout{** loaded for the language `#1'. Using the pattern for}%
\typeout{** the default language instead.}%
\else
\language=\csname l@#1\endcsname
\fi
#2}}

\bibitem{Karl00}
K.~J. Astrom, \emph{Introduction to Stochastic Control Theory}.\hskip 1em plus
  0.5em minus 0.4em\relax Courier Corporation, 1970.

\bibitem{Kothare00}
M.~V. Kothare, V.~Balakrishnan, and M.~Morari, ``Robust constrained model
  predictive control using linear matrix inequalities,'' \emph{Automatica},
  vol.~32, no.~10, pp. 1361--1379, 1996.

\bibitem{Hewing00}
\BIBentryALTinterwordspacing
L.~Hewing, K.~P. Wabersich, M.~Menner, and M.~N. Zeilinger, ``Learning-based
  model predictive control: Toward safe learning in control,'' \emph{Annual
  Review of Control, Robotics, and Autonomous Systems}, vol.~3, no.~1, pp.
  269--296, 2020. [Online]. Available:
  \url{https://doi.org/10.1146/annurev-control-090419-075625}
\BIBentrySTDinterwordspacing

\bibitem{Dean00}
S.~Dean, S.~Tu, N.~Matni, and B.~Recht, ``Safely learning to control the
  constrained linear quadratic regulator,'' in \emph{Proceedings of the 2019
  American Control Conference}.\hskip 1em plus 0.5em minus 0.4em\relax
  Philadelphia, PA, USA: IEEE, 2019.

\bibitem{bujarbaruah2018adaptive}
M.~Bujarbaruah, X.~Zhang, and F.~Borrelli, ``Adaptive mpc with chance
  constraints for fir systems,'' 2018.

\bibitem{Tanaskovic00}
M.~Tanaskovic, L.~Fagiano, R.~Smith, and M.~Morari, ``Adaptive receding horizon
  control for constrained mimo systems,'' \emph{Automatica}, vol.~50, pp.
  3019--3029, 2014.

\bibitem{Rosolia00}
U.~Rosolia and F.~Borrelli, ``Learning model predictive control for iterative
  tasks. a data-driven control framework,'' \emph{IEEE Transactions on
  Automatic Control}, vol.~63, no.~7, pp. 1883--1896, 2017.

\bibitem{Koller00}
T.~Koller, F.~Berkenkamp, M.~Turchetta, and A.~Krause, ``Learning-based model
  predictive control for safe exploration,'' \emph{arXiv}, 2018.

\bibitem{Cheng2019EndtoEndSR}
R.~Cheng, G.~Orosz, R.~Murray, and J.~Burdick, ``End-to-end safe reinforcement
  learning through barrier functions for safety-critical continuous control
  tasks,'' in \emph{AAAI}, 2019.

\bibitem{Fan2020BayesianLA}
D.~D. Fan, J.~Nguyen, R.~Thakker, N.~Alatur, A.~akbar Agha-mohammadi, and
  E.~Theodorou, ``Bayesian learning-based adaptive control for safety critical
  systems,'' \emph{2020 IEEE International Conference on Robotics and
  Automation (ICRA)}, pp. 4093--4099, 2020.

\bibitem{Choi00}
J.~Choi, F.~Castañeda, C.~J. Tomlin, and K.~Sreenath, ``Reinforcement learning
  for safety-critical control under model uncertainty, using control lyapunov
  functions and control barrier functions,'' 2020.

\bibitem{Westenbroek00}
T.~Westenbroek, A.~Agrawal, F.~Castaneda, S.~Sastry, and K.~Sreenath,
  ``Combining model-based design and model-free policy optimization to learn
  safe, stabilizing controllers,'' in \emph{Proceedings of the 7th IFAC
  Conference on Analysis and Design of Hybrid Systems}, 2021.

\bibitem{Garcia00}
J.~Garcia and F.~Fernandes, ``A comprehensive survey on safe reinforcement
  learning,'' \emph{Journal of Machine Learning Research}, vol.~16, pp.
  1437--1480, 2016.

\bibitem{NEURIPS2018_4fe51490}
Y.~Chow, O.~Nachum, E.~Duenez-Guzman, and M.~Ghavamzadeh, ``A lyapunov-based
  approach to safe reinforcement learning,'' in \emph{Advances in Neural
  Information Processing Systems}, S.~Bengio, H.~Wallach, H.~Larochelle,
  K.~Grauman, N.~Cesa-Bianchi, and R.~Garnett, Eds., vol.~31.\hskip 1em plus
  0.5em minus 0.4em\relax Curran Associates, Inc., 2018.

\bibitem{safcrit00}
\BIBentryALTinterwordspacing
K.~Srinivasan, B.~Eysenbach, S.~Ha, J.~Tan, and C.~Finn, ``Learning to be safe:
  Deep {RL} with a safety critic,'' \emph{CoRR}, vol. abs/2010.14603, 2020.
  [Online]. Available: \url{https://arxiv.org/abs/2010.14603}
\BIBentrySTDinterwordspacing

\bibitem{preexsafe0}
\BIBentryALTinterwordspacing
J.~Garc{\'{\i}}a and F.~Fern{\'{a}}ndez, ``Safe exploration of state and action
  spaces in reinforcement learning,'' \emph{CoRR}, vol. abs/1402.0560, 2014.
  [Online]. Available: \url{http://arxiv.org/abs/1402.0560}
\BIBentrySTDinterwordspacing

\bibitem{preexsafe00}
T.~J. Perkins and A.~G. Barto, ``Lyapunov design for safe reinforcement
  learning,'' \emph{J. Mach. Learn. Res.}, vol.~3, no. null, p. 803–832, mar
  2003.

\bibitem{dvp00}
\BIBentryALTinterwordspacing
S.~Levine, C.~Finn, T.~Darrell, and P.~Abbeel, ``End-to-end training of deep
  visuomotor policies,'' \emph{CoRR}, vol. abs/1504.00702, 2015. [Online].
  Available: \url{http://arxiv.org/abs/1504.00702}
\BIBentrySTDinterwordspacing

\bibitem{deepc00}
J.~Coulson, J.~Lygeros, and F.~Dorfler, ``Distributionally robust chance
  constrained data-enabled predictive control,'' \emph{IEEE Transactions on
  Automatic Control}, pp. 1--1, 2021.

\bibitem{MPC_new00}
\BIBentryALTinterwordspacing
Z.~Zhong, E.~A. del Rio-Chanona, and P.~Petsagkourakis, ``Data-driven
  distributionally robust mpc using the wasserstein metric,'' 2021. [Online].
  Available: \url{https://arxiv.org/abs/2105.08414}
\BIBentrySTDinterwordspacing

\bibitem{Nilim00}
A.~Nilim and L.~E. Ghaoui, ``Robust control of markov decision processes with
  uncertain transition matrices,'' \emph{Operations Research}, vol.~53, no.~5,
  2005.

\bibitem{Khojasteh2020ProbabilisticSC}
M.~J. Khojasteh, V.~Dhiman, M.~Franceschetti, and N.~Atanasov, ``Probabilistic
  safety constraints for learned high relative degree system dynamics,'' in
  \emph{L4DC}, 2020.

\bibitem{VanParys00}
B.~P.~V. Parys, D.~Kuhn, P.~J. Goulart, and M.~Morari, ``Distributionally
  robust control of constrained stochastic systems,'' \emph{IEEE Transactions
  on Automatic Control}, vol.~61, no.~2, pp. 430--442, 2016.

\bibitem{Paulson00}
J.~Paulson, E.~Buehler, and A.~Mesbah, ``Arbitrary polynomial chaos for
  uncertainty propagation of correlated random variables in dynamic systems,''
  \emph{IFAC PapersOnLine}, vol.~50, no.~1, pp. 3548--3553, 2017.

\bibitem{Esfahani00}
P.~Esfahani and D.~Kuhn, ``Data-driven distributionally robust optimization
  using the wasserstein metric: Performance guarantees and tractable
  reformulations,'' \emph{Mathematical Programming}, vol. 171, no. 1--2, pp.
  115--166, 2018.

\bibitem{Gao00}
R.~Gao and A.~J. Kleywegt, ``Distributionally robust stochastic optimization
  with wasserstein distance,'' \emph{arXiv}, 2016.

\bibitem{Zhao00}
C.~Zhao and Y.~Guan, ``Data-driven risk-averse stochastic optimization with
  wasserstein metric,'' \emph{Operations Research Letters}, vol.~46, no.~2, pp.
  262--267, 2018.

\bibitem{Yang00}
I.~Yang, ``Wasserstein distributionally robust stochastic control: A
  data-driven approach,'' \emph{arXiv}, 2018.

\bibitem{kandel2021distributionally}
A.~Kandel, S.~Park, and S.~Moura, ``Distributionally robust surrogate optimal
  control for high-dimensional systems,'' 2021.

\bibitem{Kandel01}
A.~Kandel and S.~Moura, ``Safe wasserstein constrained deep q-learning,''
  \emph{arXiv}, 2020.

\bibitem{NIPS2019_8942}
E.~Lecarpentier and E.~Rachelson, ``Non-stationary markov decision processes, a
  worst-case approach using model-based reinforcement learning,'' in
  \emph{Advances in Neural Information Processing Systems 32}.\hskip 1em plus
  0.5em minus 0.4em\relax Curran Associates, Inc., 2019, pp. 7216--7225.

\bibitem{asadi2018lipschitz}
K.~Asadi, D.~Misra, and M.~L. Littman, ``Lipschitz continuity in model-based
  reinforcement learning,'' 2018.

\bibitem{Akbar00}
I.~Akbar, ``Uncertainty estimation in continuous models applied to
  reinforcement learning,'' Ph.D. dissertation, UC San Diego, 2019.

\bibitem{Yang03}
I.~Yang, ``A convex optimization approach to distributionally robust markov
  decision processes with wasserstein distance,'' \emph{IEEE Control Systems
  Letters}, vol.~1, no.~1, pp. 164--169, 2017.

\bibitem{KandelGitHub}
\BIBentryALTinterwordspacing
A.~Kandel, ``{Wasserstein Nonlinear MPC},'' Aug. 2023. [Online]. Available:
  \url{https://github.com/aaronkandel/Wasserstein-Nonlinear-MPC/tree/main}
\BIBentrySTDinterwordspacing

\bibitem{Duan00}
C.~Duan, W.~Fang, L.~Jiang, L.~Yao, and J.~Liu, ``Distributionally robust
  chance-constrained approximate ac-opf with wasserstein metric,'' \emph{IEEE
  Transactions on Power Systems}, vol.~33, no.~5, pp. 4924--4936, 2018.

\bibitem{bandmbrl}
\BIBentryALTinterwordspacing
B.~Zhang, R.~Rajan, L.~Pineda, N.~O. Lambert, A.~Biedenkapp, K.~Chua,
  F.~Hutter, and R.~Calandra, ``On the importance of hyperparameter
  optimization for model-based reinforcement learning,'' \emph{CoRR}, vol.
  abs/2102.13651, 2021. [Online]. Available:
  \url{https://arxiv.org/abs/2102.13651}
\BIBentrySTDinterwordspacing

\bibitem{hyperband}
\BIBentryALTinterwordspacing
L.~Li, K.~G. Jamieson, G.~DeSalvo, A.~Rostamizadeh, and A.~Talwalkar,
  ``Efficient hyperparameter optimization and infinitely many armed bandits,''
  \emph{CoRR}, vol. abs/1603.06560, 2016. [Online]. Available:
  \url{http://arxiv.org/abs/1603.06560}
\BIBentrySTDinterwordspacing

\bibitem{Berk00}
F.~Berkenkamp, M.~Turchetta, A.~Schoellig, and A.~Krause, ``Safe model-based
  reinforcement learning with stability guarantees,'' in \emph{Advances in
  Neural Information Processing Systems}, I.~Guyon, U.~V. Luxburg, S.~Bengio,
  H.~Wallach, R.~Fergus, S.~Vishwanathan, and R.~Garnett, Eds., vol.~30.\hskip
  1em plus 0.5em minus 0.4em\relax Curran Associates, Inc., 2017.

\bibitem{lillicrap2015continuous}
T.~P. Lillicrap, J.~J. Hunt, A.~Pritzel, N.~Heess, T.~Erez, Y.~Tassa,
  D.~Silver, and D.~Wierstra, ``Continuous control with deep reinforcement
  learning,'' 2015.

\bibitem{Achiam00}
J.~Achiam, D.~Held, A.~Tamar, and P.~Abbeel, ``Constrained policy
  optimization,'' in \emph{Proceedings of the 2017 International Conference on
  Machine Learning (ICML)}.\hskip 1em plus 0.5em minus 0.4em\relax Sydney,
  Australia: PMLR, 2017.

\bibitem{Doyle00}
M.~Doyle, T.~Fuller, and J.~Newman, ``Modeling of galvanostatic charge and
  discharge of the lithium/polymer/insertion cell,'' \emph{Journal of the
  Electrochemical Society}, vol. 140, no.~6, pp. 1526--1533, 1993.

\bibitem{Perez05}
H.~Perez, X.~Hu, S.~Dey, and S.~Moura, ``Optimal charging of li-ion batteries
  with coupled electro-thermal-aging dynamics,'' \emph{IEEE Transactions on
  Vehicular Technology}, vol.~66, no.~7, pp. 7761--7770, 2017.

\bibitem{visuo}
\BIBentryALTinterwordspacing
S.~Levine, C.~Finn, T.~Darrell, and P.~Abbeel, ``End-to-end training of deep
  visuomotor policies,'' \emph{CoRR}, vol. abs/1504.00702, 2015. [Online].
  Available: \url{http://arxiv.org/abs/1504.00702}
\BIBentrySTDinterwordspacing

\bibitem{Beyer2002}
H.~Beyer and H.~Schwefel, ``Evolution strategies - a comprehensive
  introduction,'' \emph{Natural Computing}, vol.~1, no.~1, pp. 3--52, March
  2002.

\bibitem{RStest}
\BIBentryALTinterwordspacing
H.~Mania, A.~Guy, and B.~Recht, ``Simple random search provides a competitive
  approach to reinforcement learning,'' \emph{CoRR}, vol. abs/1803.07055, 2018.
  [Online]. Available: \url{http://arxiv.org/abs/1803.07055}
\BIBentrySTDinterwordspacing

\end{thebibliography}

\begin{figure*}[t]
      \centering  
      \includegraphics[trim = 0mm 0mm 0mm 0mm, clip, width=\textwidth]{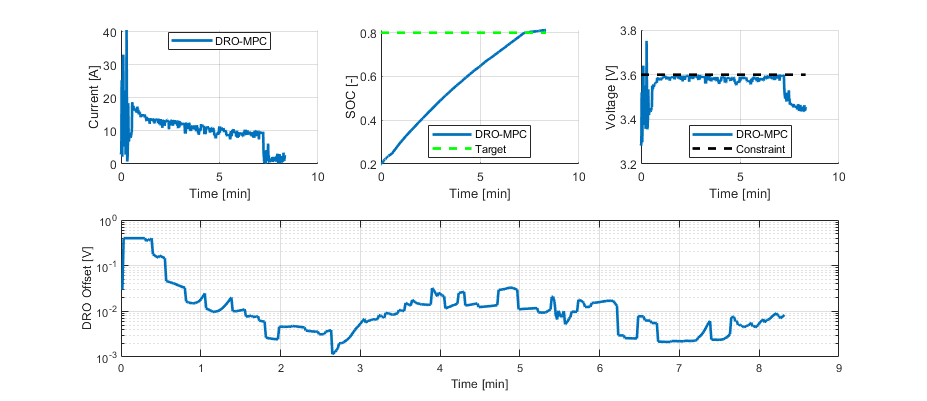}
      \caption{Battery experiment showing time evolution of the DRO offset and added PoE component. The PoE component adds noise to the input signal while maintaining probabilistic feasibility. We cap the DRO offset at $r_{DRO;max} = 0.4$ (the max true value was temporarily 14.24), which would create an empty feasible set. Remark 8 in Section IV.E describes how implementation works when the DRO feasible set is ostensibly empty. } 
      \label{figurelabel8}
\end{figure*}

\section*{Appendix}

\subsection*{Cardinality of Constraints Remains Constant}
In the following lemma, we prove that the number of constraints in the reformulation of the DRO problem in (\ref{}) need only be $m$, where $g(\cdot) \in \mathbb{R}^n \times \mathbb{R} \rightarrow \mathbb{R}^m$. When g() is non-separable, as described in \cite{Duan00}, then the number of constraints in the reformulation scales super-linearly as  $2^m$.

\begin{lemma}
    If the modeling error residuals are defined using the relation:
    \begin{subequations}
    \begin{align}
    R_1^{(t)} &= |g(x_t, u_t^*) - \hat{g}(x_t, u_t^*, \theta_g)| \\
    R_1^{(t+1)} &= |g(x_{t+1}^*, u_{t+1}^*) - \hat{g}(\hat{x}_{t+1}, u_{t+1}^*, \theta_g)| \\
    R_1^{(t+2)} &= |g(x_{t+2}^*, u_{t+2}^*) - \hat{g}(\hat{x}_{t+2}, u_{t+2}^*, \theta_g)| \\
    \vdots  \: &
    \end{align}   
    \end{subequations}
and appear in the constraint function $g(\cdot)$ as (\ref{lmpcfullcon}), then the number of constraints in the reformulated problem remains identically $m$ without jeopardizing the probabilistic guarantee.
\end{lemma}

\begin{proof}
    Consider the following stochastic constraint converted to a distributionally robust chance constraint:
    \begin{subequations}
    \begin{align}
        x+\textbf{R} &\leq 0\\
        \hat{\mathbb{P}} \left[ x+\textbf{R} \leq 0 \right] &\geq 1 - \eta\\
        \underset{\mathbb{P} \in \mathbb{B}_\epsilon} {\text{inf}} \mathbb{P} \left[ x+\textbf{R} \leq 0 \right] &\geq 1 - \eta
    \end{align}   
    \end{subequations}
    representing a constraint with uncertainty.  Without loss of generality, we consider a learning MPC program with horizon $N=1$.
    
    The method of \cite{Duan00} enumerates across the vertices of a hypercube by modulating the sign of the DRO variable $\sigma$. However, when the random variable is a separable offset from a constant constraint boundary, we only need consider perturbations that add conservatism. In the 1-dimensional case, we can see from looking at the set of constraints
    \begin{equation}
        x \leq -r \: \text{and} \: x \leq r
    \end{equation}
    that only the first constraint $x \leq -r$ will ever be active. Therefore, $x \leq -r$ adequately defines the feasible region. 

    Likewise, if we consider the case where $\textbf{R}\in \mathbb{R}^2$ with additive $\textbf{R}$, we obtain the following set of constraints
    \begin{align}
        \begin{bmatrix}
            \tilde{x}_1\\
            \tilde{x}_2
        \end{bmatrix} + \begin{bmatrix}
            r_1\\
            r_2
        \end{bmatrix} &\leq 0. \label{con1}\\
        \begin{bmatrix}
            \tilde{x}_1\\
            \tilde{x}_2
        \end{bmatrix} + \begin{bmatrix}
            -r_1\\
            r_2
        \end{bmatrix} &\leq 0 \label{con2}\\
        \begin{bmatrix}
            \tilde{x}_1\\
            \tilde{x}_2
        \end{bmatrix} + \begin{bmatrix}
            r_1\\
            -r_2
        \end{bmatrix} &\leq 0\\
        \begin{bmatrix}
            \tilde{x}_1\\
            \tilde{x}_2
        \end{bmatrix} + \begin{bmatrix}
            -r_1\\
            -r_2
        \end{bmatrix} &\leq 0 \label{con4}
    \end{align}
    we see trivially that the feasible region defined by (\ref{con1}-\ref{con4}) is identical to that defined solely by (\ref{con4}). This pattern continues for any $m \in \mathbb{N}$ of $\textbf{R} \in \mathbb{R}^m$.

\end{proof}

\subsection*{Evolutionary Strategies and Random Search}

In our paper, we utilize a $(1+\lambda)$ evolutionary strategy to approximately solve the numerical MPC optimization program. This is a form of random search, where instead of utilizing gradients for optimization, we utilize a random strategy to iteratively test mutations of our initial guess until converging to a reasonable approximately optimal solution. 

This is a subset of what is generally referred to as a $(\frac{\mu}{\rho}+\lambda)$ evolutionary strategy, whose precise definition can be referenced in \cite{Beyer2002}. A $(\frac{\mu}{\rho}+\lambda)$ evolutionary strategy is a very simple form of a genetic algorithm, whereby at each generation/iteration of optimization, we have some number of ``parents'' who are mutated, and the parents are replaced by the highest performing mutated offspring. Random search has been shown to be a highly effective method for solving optimization problems in reinforcement learning literature \cite{RStest}. Random search is also highly amenable to constrained optimization (without equality constraints), as infeasible mutants can be pruned from selection.  Furthermore, if no feasible mutants are found, the mutant that least violates the constraint boundary can be defaulted to if additional computation is undesirable.

\subsection*{Slow Model Adaptation}
To accommodate potential cases where the true plant dynamics change slowly over time, we can adopt the following approach which preserves the safety guarantees of the Wasserstein DRO framework.  We have system dynamics $x \in \mathbb{R}^n$ with no finite escape time. Furthermore, $g(x, u, \theta^*) \leq 0$ is our constraint function. Suppose it holds that the function $g$ behaves in the following manner (similarly, although not identically, to a Lipschitz continuous function):
\begin{equation}
    \underset{x \in \mathcal{x},u \in \mathcal{U}, \delta \theta}{\text{max}} | {g}(x, u, \theta + \delta \theta) - g(x, u, \theta) | \: \leq C 
\end{equation}
where $\delta \theta = \theta^*_{t+1}-\theta^*_t$ is any possible deviation in the model parameters over the course of a single timestep. The value $\delta \theta$ is bounded.  Consider we are at time $t$ of the experiment.  Let us represent the 1-step residual at time $j = t - k$, where $k \in \{1, 2, ..., t\}$ is an integer, as:
\begin{equation}
    R_1^{(t)} = g(x_t, u_t, \theta^*_t) -  \hat{g}(x_t, u_t, \theta_t)
\end{equation}
where $\theta^*_t$ is the parameterization of the true plant at time $t$, and $\theta_t$ is the learned model at time $t$.
If we add a value to the residual $R_1^{(t)}$ of $C \cdot k \cdot \text{sgn}(R_1^{(t)})$,
\begin{equation}
    \tilde{R}_1^{(t)} = R_1^{(t)} + C \cdot k \cdot \text{sgn}(R_1^{(t)})
\end{equation}
we accommodate for worst-case model adaptation in our algorithm. This scheme, coupled with a judiciously designed moving window of residuals, can accommodate model adaptation in the true underlying plant.

This provides a conservative, but robust means to address additional model adaptation throughout the learning process. Ideally the bound on the change of the constraint function $C$ is small, meaning the true plant changes gradually over time.  In this case, the additional offset will present a relatively small additional contribution to the overall robust offset.

\subsection*{Visualization of DRO Offset and PoE Demonstration}

To visualize both an added PoE component and the DRO offset, we run the following additional experiment, plotting the evolution of the offset throughout time. Here, we consider a set $\mathcal{N}$ of additive Uniform noise to the control input capped at $\pm 5$.  Figure 7 shows these results.



\end{document}